\documentclass[letterpaper,twocolumn,10pt]{article} \usepackage{usenix,epsfig,endnotes}
\usepackage[utf8]{inputenc}
\usepackage[cmex10]{amsmath}
\usepackage{array}
\usepackage[hyphens]{url}
\usepackage[table]{xcolor}
\usepackage{xspace}
\PassOptionsToPackage{hyphens}{url}\usepackage{url}
\usepackage{tikz}
\usepackage{multirow}
\usepackage{pgfplots}
\usetikzlibrary{pgfplots.groupplots}
\usetikzlibrary{external}
\usetikzlibrary{arrows}
\usetikzlibrary{positioning}
\usetikzlibrary{decorations.pathreplacing}
\usetikzlibrary{shapes.arrows}
\usetikzlibrary{pgfplots.groupplots}
\usepackage{letltxmacro}
\usepackage{cite} 
\pgfplotsset{compat=newest}
\usepackage{ifthenx}
\usepackage{collcell}
\usepackage{nicefrac}
\usepackage{amsmath}
\usepackage[all]{nowidow}
\usepackage{filecontents}
\usepackage{listings}
\usepackage{lipsum}
\usepackage{todonotes}
\usepackage{makecell}
\usepackage{paralist}
\usepackage{flushend}
\usepackage{ifthenx}
\usepackage{makecell}
\usepackage{paralist}
\usepackage{xspace}
\usepackage{soul} 
\usepackage{csvenhanced}
\usepackage{etex}
\usepackage{bytefield}
\usepackage{amsfonts}
\usepackage{paralist}
\usepackage{tcolorbox}
\usepackage{arydshln}
\usepackage{siunitx,array,booktabs}
\usepackage{enumitem}

\LetLtxMacro{\oldtodo}{\todo}
\renewcommand{\todo}[2][]{\tikzexternaldisable\oldtodo[fancyline,size=\footnotesize,#1]{#2}\tikzexternalenable}
\renewcommand{\todo}[1]{\tikzexternaldisable\oldtodo[fancyline,size=\footnotesize]{#1}\tikzexternalenable}

\usepackage{algorithmic}
\usepackage[ruled]{algorithm2e}

\usepackage{color}
\usepackage{paralist}
\lstdefinestyle{mystyle}{
    commentstyle=\color{green!40!black},
    basicstyle=\footnotesize,
    breakatwhitespace=false,
    breaklines=true,
    captionpos=b,
    keepspaces=true,
    numbers=left,
    numbersep=5pt,
    showspaces=false,
    showstringspaces=false,
    showtabs=false,
    tabsize=2,
    xleftmargin=15pt
}

\newcommand{\etal}{et~al.\ } 
\newcommand{\ie}{\textit{i.e.},\ } 
\newcommand{\eg}{e.g.,\ } 
\newcommand{\cf}{cf.\ } 

\newcommand{\FlushOnly}{\emph{Flush+Flush}\xspace}

\newcommand{\FlushReload}{\emph{Flush+Reload}\xspace}
\newcommand{\EvictReload}{\emph{Evict+Reload}\xspace}
\newcommand{\EvictTime}{\emph{Evict+Time}\xspace}
\newcommand{\PrimeProbe}{\emph{Prime+Probe}\xspace}
\newcommand{\Alcatel}{Alcatel One Touch Pop 2\xspace}
\newcommand{\OnePlus}{OnePlus One\xspace}
\newcommand{\Samsung}{Samsung Galaxy S6\xspace}
\mathchardef\mhyphen="2D

\newif\ifanonymous
 \anonymoustrue

\begin{document}
\title{ARMageddon: Cache Attacks on Mobile Devices} 

\author{
{\rm Moritz Lipp}, {\rm Daniel Gruss}, {\rm Raphael Spreitzer}, {\rm Clémentine Maurice}, and {\rm Stefan Mangard}\\
Graz University of Technology, Austria 
}

\maketitle

\begin{abstract}
In the last 10 years, cache attacks on Intel x86 CPUs have gained increasing attention among the scientific community and powerful techniques to exploit cache side channels have been developed. However, modern smartphones use one or more multi-core ARM CPUs that have a different cache organization and instruction set than \mbox{Intel} x86 CPUs.
So far, no cross-core cache attacks have been demonstrated on non-rooted Android smartphones.
In this work, we demonstrate how to solve key challenges to perform the most powerful cross-core cache attacks
\PrimeProbe, \FlushReload, \EvictReload, and \FlushOnly on non-rooted ARM-based devices without any privileges. 
Based on our techniques, we demonstrate covert channels that outperform state-of-the-art covert channels on Android by several orders of magnitude.
Moreover, we present attacks to monitor tap and swipe events as well as keystrokes, and even derive the lengths of words entered on the touchscreen. 
Eventually, we are the first to attack cryptographic primitives implemented in Java. Our attacks work across CPUs and can even monitor cache activity in the ARM TrustZone from the normal world. 
The techniques we present can be used to attack hundreds of millions of Android devices. 
\end{abstract}

\section{Introduction}\label{sec:intro}
\bgroup
\let\thefootnote\relax\footnotetext{Original publication in the Proceedings of the 25th Annual USENIX Security Symposium (USENIX Security 2016).\\\url{https://www.usenix.org/conference/usenixsecurity16/technical-sessions/presentation/lipp}}
\egroup

Cache attacks represent a powerful means of exploiting the different access times within the memory hierarchy of modern system architectures. Until recently, these attacks explicitly targeted cryptographic implementations, for instance, by means of cache timing attacks~\cite{2004-bernstein-cachetiming} or the well-known \EvictTime and \PrimeProbe techniques~\cite{DBLP:conf/ctrsa/OsvikST06}. The seminal paper by Yarom and Falkner~\cite{DBLP:conf/uss/YaromF14} introduced the so-called \FlushReload attack, which allows an attacker to infer which specific parts of a binary are accessed by a victim program with an unprecedented accuracy and probing frequency. 
Recently, Gruss~\etal\cite{DBLP:conf/uss/GrussSM15} demonstrated the possibility to use \FlushReload to automatically exploit cache-based side channels via cache template attacks on Intel platforms. \FlushReload does not only allow for efficient attacks against cryptographic implementations~\cite{DBLP:conf/ches/BengerPSY14,DBLP:conf/raid/ApececheaIES14,DBLP:conf/ctrsa/PolSY15}, but also to infer keystroke information and even to build keyloggers on Intel platforms~\cite{DBLP:conf/uss/GrussSM15}. In contrast to attacks on cryptographic algorithms, which are typically triggered multiple times, these attacks require a significantly higher accuracy as an attacker has only one single chance to observe a user input event.

Although a few publications about cache attacks on AES T-table implementations on mobile devices exist~\cite{DBLP:conf/ctrsa/BogdanovEPW10,DBLP:conf/fc/WeissHS12,DBLP:conf/nss/SpreitzerP13,DBLP:conf/cosade/SpreitzerP13,DBLP:conf/wistp/SpreitzerG14},
the more efficient cross-core attack techniques \PrimeProbe, \FlushReload, \EvictReload, and \FlushOnly~\cite{DBLP:journals/corr/GrussMW15} have not been applied on smartphones. In fact, there was reasonable doubt~\cite{DBLP:conf/uss/YaromF14} whether these cross-core attacks can be mounted on ARM-based devices at all.
In this work, we demonstrate that these attack techniques are applicable on ARM-based devices by solving the following key challenges systematically:
\begin{compactenum}
\item \textit{Last-level caches are not inclusive on ARM and thus cross-core attacks cannot rely on this property.} 
Indeed, existing cross-core attacks exploit the inclusiveness of shared last-level caches~\cite{DBLP:conf/sp/LiuYGHL15,DBLP:conf/dimva/MauriceNHF15,DBLP:conf/raid/MauriceSNHF15,DBLP:conf/ccs/OrenKSK15,DBLP:conf/uss/GrussSM15,DBLP:conf/uss/YaromF14,DBLP:conf/sp/IrazoquiES15,DBLP:journals/corr/GrussMW15,DBLP:conf/cosade/GulmezogluIAES15} and, thus, 
no cross-core attacks have been demonstrated on ARM so far. We present an approach that exploits coherence protocols and L1-to-L2 transfers to make these attacks applicable on mobile devices with non-inclusive shared last-level caches, irrespective of the cache organization.\footnote{Simultaneously to our work on ARM, Irazoqui~\etal\cite{DBLP:conf/ccs/IrazoquiES16} developed a technique to exploit cache coherence protocols on AMD x86 CPUs and mounted the first cross-CPU cache attack.} 
\item \textit{Most modern smartphones have multiple CPUs that do not share a cache.} However, cache coherence protocols allow CPUs to fetch cache lines from remote cores faster than from the main memory. We utilize this property to mount both cross-core and cross-CPU attacks.
\item \textit{Except ARMv8-A CPUs, ARM processors do not support a flush instruction.} In these cases, a fast eviction strategy must be applied for 
high-frequency measurements. 
As existing eviction strategies are too slow, we analyze more than 4\,200 eviction strategies for our test devices, based on Rowhammer attack techniques~\cite{rowhammerjs}. 
\item \textit{ARM CPUs use a pseudo-random replacement policy} to decide which cache line to replace within a cache set. This introduces additional noise even for robust time-driven cache attacks~\cite{DBLP:conf/nss/SpreitzerP13,DBLP:conf/wistp/SpreitzerG14}. For the same reason, \PrimeProbe has been an open challenge~\cite{DBLP:conf/cosade/SpreitzerP13} on ARM, as an attacker needs to predict which cache line will be replaced first and wrong predictions destroy measurements. We design re-access loops that interlock with a cache eviction strategy to reduce the effect of 
wrong predictions. 
\item \textit{Cycle-accurate timings require root access on ARM}~\cite{arm_arch_manualv7} and alternatives have not been evaluated so far.  
We evaluate different timing sources and show that cache attacks can be mounted in any case.
\end{compactenum}

Based on these building blocks, we demonstrate practical and highly efficient cache attacks on ARM.\footnote{Source code for ARMageddon attack examples can be found at \url{https://github.com/IAIK/armageddon}.} We do not restrict our investigations to cryptographic implementations but also consider cache attacks as a means to infer other sensitive information---such as inter-keystroke timings or the length of a swipe action---requiring a significantly higher measurement accuracy. 
Besides these generic attacks, we also demonstrate that cache attacks can be used to monitor cache activity caused within the ARM TrustZone from the normal world.
Nevertheless, we do not aim to exhaustively list possible exploits or find new attack vectors on cryptographic algorithms. Instead, we aim to demonstrate the immense attack potential of the presented cross-core and cross-CPU attacks on ARM-based mobile devices based on well-studied attack vectors. 
Our work allows to apply existing attacks to millions of off-the-shelf Android devices without any privileges. 
Furthermore, our investigations show that Android still employs vulnerable AES T-table implementations.

\paragraph{Contributions.} 
The contributions of this work are:
\begin{compactitem}
\item We demonstrate the applicability of highly efficient cache attacks like \PrimeProbe, \FlushReload, \EvictReload, and \FlushOnly on ARM. 

\item Our attacks work irrespective of the actual cache organization and, thus, are the first last-level 
cache attacks that can be applied cross-core and also 
cross-CPU on off-the-shelf ARM-based devices. More specifically, our attacks work against last-level caches that are instruction-inclusive and data-non-inclusive as well as caches that are instruction-non-inclusive and data-inclusive.

\item Our cache-based covert channel outperforms all existing covert channels on Android by several orders of magnitude.

\item We demonstrate the power of these attacks by attacking cryptographic implementations and by inferring more fine-grained information like keystrokes and swipe actions on the touchscreen. 
\end{compactitem}

\paragraph{Outline.} 
The remainder of this paper is structured as follows. In Section~\ref{sec:background}, we provide information on background and related work. Section~\ref{sec:attack} describes the techniques that are the building blocks for our attacks. In Section~\ref{sec:covert}, we demonstrate and evaluate fast cross-core and cross-CPU covert channels on Android. In Section~\ref{sec:userinput}, we demonstrate cache template attacks on user input events. In Section~\ref{sec:crypto}, we present attacks on cryptographic implementations used in practice as well the possibility to observe cache activity of cryptographic computations within the TrustZone. We discuss countermeasures in Section~\ref{sec:countermeasures} and conclude this work in Section~\ref{sec:conclusion}.

\section{Background and Related Work}\label{sec:background}
In this section, we provide the required preliminaries and discuss related work in the context of cache attacks. 

\subsection{CPU Caches}\label{sec:caches}
Today's CPU performance is influenced not only by the clock frequency but also by the latency of instructions, operand
fetches, and other interactions with internal and external devices. In order to overcome the latency of system memory accesses, CPUs employ caches to buffer frequently used data in small and fast internal memories. 

Modern caches organize cache lines in multiple sets, which is also known as set-associative caches. Each memory address maps to one of these cache sets and addresses that map to the same cache set are considered congruent.
Congruent addresses compete for cache lines within the same set and a predefined replacement policy determines which cache line is replaced. 
For instance, the last generations of Intel CPUs employ an undocumented variant of least-recently used (LRU) replacement policy~\cite{rowhammerjs}. ARM processors use a pseudo-LRU replacement policy for the L1 cache and they support two different cache replacement policies for L2 caches, namely round-robin and pseudo-random replacement policy. In practice, however, only the pseudo-random replacement policy is used due to performance reasons. Switching the cache replacement policy is only possible in privileged mode. The implementation details for the pseudo-random policy are not documented. 

CPU caches can either be virtually indexed or physically indexed, which determines whether 
the index is derived from the virtual or physical address.
A so-called tag 
uniquely identifies the address that is cached within a specific cache line.
Although this tag can also be based on the virtual or physical address, most modern caches use physical tags because they can be computed simultaneously while locating the cache set. ARM 
typically uses physically indexed, physically tagged L2 caches. 

CPUs have multiple cache levels, with the lower levels being faster and smaller than the higher levels. ARM processors typically have two levels of cache. 
If all cache lines from lower levels are also stored in a higher-level cache, the higher-level cache is called \textit{inclusive}.
If a cache line can only reside in one of the cache levels at any point in time, the caches are called \textit{exclusive}.
If the cache is neither inclusive nor exclusive, it is called \textit{non-inclusive}.
The last-level cache is often shared among all cores to enhance the performance upon transitioning threads between cores
and to simplify cross-core cache lookups. However, with shared last-level caches, one core
can (intentionally) influence the cache content of all other cores. This represents the basis for cache attacks like \FlushReload~\cite{DBLP:conf/uss/YaromF14}.

In order to keep caches of multiple CPU cores or CPUs in a coherent state, so-called coherence protocols are employed. However, coherence protocols also introduce exploitable timing effects, which has recently been exploited by Irazoqui~\etal\cite{DBLP:conf/ccs/IrazoquiES16} on x86 CPUs.

\begin{table*}
  \centering
  \caption{Test devices used in this paper.}
  \vspace{5pt}
\resizebox{.999\hsize}{!}{
  \begin{tabular}{p{0.11\hsize}p{0.15\hsize}p{0.135\hsize}p{0.135\hsize}p{0.16\hsize}p{0.22\hsize}}
  \toprule 
Device & SoC & CPU (cores) & L1 caches & L2 cache & Inclusiveness \\ 
  \midrule
  \OnePlus & Qualcomm \newline Snapdragon 801 & Krait 400 (2) \newline 2.5\,GHz & 2$\times$ 16\,KB, \newline 4-way, 64 sets & 2\,048\,KB, \newline 8-way, 2\,048 sets & non-inclusive \\
\hdashline
  \Alcatel & Qualcomm \newline Snapdragon 410 & Cortex-A53 (4) \newline 1.2\,GHz & 4$\times$ 32\,KB, \newline 4-way, 128 sets & 512\,KB, \newline 16-way, 512 sets & instruction-inclusive, \newline data-non-inclusive \\
\hdashline
             &                & Cortex-A53 (4) & 4$\times$ 32\,KB, & 256\,KB,            & instruction-inclusive, \\
   Samsung   & Samsung Exynos & 1.5\,GHz       & 4-way, 128 sets   & 16-way, 256 sets    & data-non-inclusive \\
   Galaxy S6 & 7 Octa 7420    & Cortex-A57 (4) & 4$\times$ 32\,KB, & 2\,048\,KB,         & instruction-non-inclusive, \\ 
             &                & 2.1\,GHz       & 2-way, 256 sets   & 16-way, 2\,048 sets & data-inclusive \\
  \bottomrule
 \end{tabular}
}
  \label{tbl:test_devices}
\end{table*}

In this paper, we demonstrate attacks on three smartphones as listed in Table~\ref{tbl:test_devices}.
The Krait 400 is an ARMv7-A CPU, the other two processors are ARMv8-A CPUs. However, the stock Android of the \Alcatel is compiled for an ARMv7-A instruction set and thus ARMv8-A instructions are not used. We generically refer to ARMv7-A and ARMv8-A as ``ARM architecture'' throughout this paper. All devices have a shared L2 cache. On the \Samsung, the flush instruction is unlocked by default, which means that it is available in userspace. Furthermore, all devices employ a cache coherence protocol between cores and on the \Samsung even between the two CPUs~\cite{a53_manual}.

\subsection{Shared Memory}\label{shm}
Read-only shared memory can be used as a means of memory usage optimization.
In case of shared libraries it reduces the memory footprint and enhances the speed by lowering cache contention. The operating system implements this behavior by mapping the same physical memory into the address space of each process. As this memory sharing mechanism is independent of how a file was opened or accessed, an attacker can map a binary to have read-only shared memory with a victim program. A similar effect is caused by content-based page deduplication where physical pages with identical content are merged.

Android applications are usually written in Java and, thus, contain self-modifying code or just-in-time compiled code. This code would typically not be shared. Since Android version 4.4 the Dalvik VM was gradually replaced by the Android Runtime (ART). With ART, Java byte code is compiled to native code binaries~\cite{art} and thus can be shared too.

\subsection{Cache Attacks}\label{cacheattacks}
Initially, cache timing attacks were performed on cryptographic algorithms~\cite{DBLP:conf/crypto/Kocher96,DBLP:journals/jcs/KelseySWH00,DBLP:journals/iacr/Page02,DBLP:conf/ches/TsunooSSSM03,2004-bernstein-cachetiming,Neve2006CachebasedVulnerabilities,DBLP:conf/ccs/NeveSW06}. For example, Bernstein~\cite{2004-bernstein-cachetiming} 
exploited the total execution time of AES T-table implementations. More fine-grained exploitations of memory accesses to the CPU cache have been proposed by Percival~\cite{2005-percival-cache}
and Osvik~\etal\cite{DBLP:conf/ctrsa/OsvikST06}. More specifically, Osvik~\etal formalized two concepts, 
namely \EvictTime and \PrimeProbe, to determine which specific cache sets were accessed by a victim program. Both 
approaches consist of three basic steps. 

\noindent\textbf{\EvictTime:} 
\begin{compactenum}
 \item Measure execution time of victim program. 
 \item Evict a specific cache set. 
 \item Measure execution time of victim program again. 
\end{compactenum}

\noindent\textbf{\PrimeProbe:}
\begin{compactenum}
 \item Occupy specific cache sets.
 \item Victim program is scheduled. 
 \item Determine which cache sets are still occupied. 
\end{compactenum}

Both approaches allow an adversary to determine which cache sets 
are used during the victim's computations and have been exploited to attack cryptographic 
implementations~\cite{DBLP:conf/ctrsa/OsvikST06,DBLP:journals/joc/TromerOS10,DBLP:conf/sp/IrazoquiES15,DBLP:conf/sp/LiuYGHL15} 
and to build cross-VM covert channels~\cite{DBLP:conf/dimva/MauriceNHF15}.
Yarom and Falkner~\cite{DBLP:conf/uss/YaromF14} proposed \FlushReload, a significantly more fine-grained
attack that exploits three fundamental concepts of modern system architectures. First, the availability of shared memory between
the victim process and the adversary. Second, last-level caches are typically shared among all
cores. Third, Intel platforms use inclusive last-level caches, meaning that the eviction of information from 
the last-level cache leads to the eviction of this data from all lower-level caches of other cores, which allows
any program to evict data from other programs on other cores. While the basic idea of this attack has been proposed
by Gullasch~\etal\cite{DBLP:conf/sp/GullaschBK11}, Yarom and Falkner extended this idea to shared last-level caches, allowing cross-core attacks. \FlushReload works as follows. 

\noindent\textbf{\FlushReload:}
\begin{compactenum}
 \item Map binary (\eg shared object) 
 into address space. 
 \item Flush a cache line (code or data) from the cache.
 \item Schedule the victim program.
 \item Check if the corresponding line from step 2 has been loaded by the victim program. 
\end{compactenum}

Thereby, \FlushReload allows an attacker to determine which specific instructions are executed
and also which specific data is accessed by the victim program. Thus, rather fine-grained attacks
are possible and have already been demonstrated against cryptographic implementations~\cite{Irazoqui2015,DBLP:conf/ccs/ApececheaIES15,DBLP:conf/cosade/GulmezogluIAES15}. 
Furthermore, Gruss~\etal\cite{DBLP:conf/uss/GrussSM15} demonstrated 
the possibility to automatically exploit cache-based side-channel information based on the \FlushReload approach.
Besides attacking cryptographic implementations like AES T-table implementations, they showed how 
to infer keystroke information and even how to build a keylogger by exploiting the cache side channel. 
Similarly, Oren~\etal\cite{DBLP:conf/ccs/OrenKSK15} demonstrated the possibility to exploit
cache attacks on Intel platforms from JavaScript and showed how to infer visited websites and how to track 
the user's mouse activity.

Gruss~\etal\cite{DBLP:conf/uss/GrussSM15} proposed the \EvictReload technique that replaces the flush instruction in \FlushReload by eviction.
While it has no practical application on x86 CPUs, we show that it can be used 
on ARM CPUs.
Recently, \FlushOnly~\cite{DBLP:journals/corr/GrussMW15} has been proposed. Unlike other techniques, it does not perform any memory access but relies on the timing of the flush instruction to determine whether a line has been loaded by a victim. We show that the execution time of the ARMv8-A flush instruction also depends on whether or not data is cached and, thus, can be used to implement this attack.

While the attacks discussed above have been proposed and investigated for Intel processors, the same attacks were considered not applicable to modern smartphones due to differences in the instruction set, the cache organization~\cite{DBLP:conf/uss/YaromF14}, and in the multi-core and multi-CPU architecture. 
Thus, only same-core cache attacks have been demonstrated on smartphones so far. For instance, 
Wei\ss~\etal\cite{DBLP:conf/fc/WeissHS12} investigated \mbox{Bernstein}'s cache-timing attack~\cite{2004-bernstein-cachetiming} on a Beagleboard
employing an ARM Cortex-A8 processor. Later on, Wei\ss~\etal\cite{DBLP:conf/intrust/WeissWAS14} investigated this timing attack in a multi-core setting on a development board. 
As Wei\ss~\etal\cite{DBLP:conf/fc/WeissHS12} claimed that noise makes the attack difficult, 
Spreitzer and Plos~\cite{DBLP:conf/nss/SpreitzerP13} investigated the applicability of Bernstein's
cache-timing attack on different ARM Cortex-A8 and ARM Cortex-A9 smartphones running Android. 
Both investigations~\cite{DBLP:conf/fc/WeissHS12,DBLP:conf/nss/SpreitzerP13} confirmed
that timing information is leaking, but the attack takes several hours due to the high number of measurement samples that are required,
\ie about $2^{30}$ AES encryptions.
Later on, Spreitzer and G\'{e}rard~\cite{DBLP:conf/wistp/SpreitzerG14} improved upon these results and 
managed to reduce the key space to a complexity which is practically relevant. 

Besides Bernstein's attack, another attack against AES T-table implementations has been 
proposed by Bogdanov~\etal\cite{DBLP:conf/ctrsa/BogdanovEPW10}, who exploited so-called wide collisions
on an ARM9 microprocessor. In addition, power analysis attacks~\cite{DBLP:conf/wisa/GallaisKT10} 
and electromagnetic emanations~\cite{Gallais2011ErrorTolerancein} 
have been 
used to visualize cache accesses during AES computations on ARM microprocessors. 
Furthermore, Spreitzer and Plos~\cite{DBLP:conf/cosade/SpreitzerP13} implemented
\EvictTime~\cite{DBLP:conf/ctrsa/OsvikST06} 
in order to attack an AES T-table implementation
on Android-based smartphones. However, so far only cache attacks against AES T-table implementations
have been considered on smartphone platforms and none of the recent advances
have been demonstrated on mobile devices. 

\section{ARMageddon Attack Techniques}\label{sec:attack}
We consider a scenario where an adversary attacks a smartphone user by means of a malicious application. This application \textit{does not require any permission} and, most importantly, it can be executed in unprivileged userspace and \textit{does not require a rooted device}. As 
our attack techniques do not exploit specific vulnerabilities of Android
versions, they work on stock Android ROMs as well as customized ROMs in use
today.

\subsection{Defeating the Cache Organization}\label{sec:shared_cache_lines}
In this section, we tackle the aforementioned challenges 1 and 2, \ie the last-level cache is not inclusive and multiple processors do not necessarily share a cache level. 

When it comes to caches, ARM CPUs are very heterogeneous compared to Intel CPUs. For example, whether or not a CPU has a second-level cache can be decided by the manufacturer. 
Nevertheless, the last-level cache on ARM devices is usually shared among all cores and it can have different inclusiveness properties for instructions and data. Due to cache coherence, shared memory is kept in a coherent state across cores and CPUs. This is of importance when measuring timing differences between cache accesses and memory accesses (cache misses), as fast remote-cache accesses are performed instead of slow memory accesses~\cite{a53_manual}. In case of a non-coherent cache, a cross-core attack is not possible but an attacker can run the spy process on all cores simultaneously and thus fall back to a same-core attack. However, we observed that caches are coherent on all our test devices. 

\begin{figure}[t]
  \centering
  \resizebox {0.58\hsize} {!} {
  	\begin{tikzpicture}[scale=1,transform shape]
	\draw[black!30,fill=black!10] (0,-0.25) rectangle +(2.25,0.5) node[color=black!80,midway] {\footnotesize{Core 0}};
	\draw[black!30,fill=black!10] (0,-0.5) rectangle +(1,-0.5) node[color=black!80,midway] {\footnotesize{L1I}};
	\draw[black!30] (0.5,-0.25) -- +(0,-0.25);
	\draw[black!30] [decorate,decoration={brace,amplitude=5pt}] (0,-1.75) -- (0,-1) node[color=black!80,yshift=+5pt,midway,sloped,above] {\footnotesize{Sets}};
	\draw[black!30] (0,-1) rectangle +(0.25,-0.25);
	\draw[black!30] (0.25,-1) rectangle +(0.25,-0.25);
	\draw[black!30] (0.5,-1) rectangle +(0.25,-0.25);
	\draw[black!30] (0.75,-1) rectangle +(0.25,-0.25);
	\draw[black!30] (0,-1.25) rectangle +(0.25,-0.25);
	\draw[black!30] (0.25,-1.25) rectangle +(0.25,-0.25);

	\draw[black!30,fill=red] (0.5,-1.25) rectangle +(0.25,-0.25);
	\draw[black!30] (0.75,-1.25) rectangle +(0.25,-0.25);
	\draw[black!30] (0,-1.5) rectangle +(0.25,-0.25);
	\draw[black!30] (0.25,-1.5) rectangle +(0.25,-0.25);
	\draw[black!30] (0.5,-1.5) rectangle +(0.25,-0.25);
	\draw[black!30] (0.75,-1.5) rectangle +(0.25,-0.25);

	\draw[black!30] (1.75,-0.25) -- +(0,-0.25);
	\draw[black!30,fill=black!10] (1.25,-0.5) rectangle +(1,-0.5) node[color=black!80,midway] {\footnotesize L1D};
	\draw[black!30] (1.25,-1) rectangle +(0.25,-0.25);
	\draw[black!30] (1.5,-1) rectangle +(0.25,-0.25);
	\draw[black!30] (1.75,-1) rectangle +(0.25,-0.25);
	\draw[black!30] (2,-1) rectangle +(0.25,-0.25);
	\draw[black!30] (1.25,-1.25) rectangle +(0.25,-0.25);
	\draw[black!30] (1.5,-1.25) rectangle +(0.25,-0.25);
	\draw[black!30] (1.75,-1.25) rectangle +(0.25,-0.25);
	\draw[black!30] (2,-1.25) rectangle +(0.25,-0.25);
	\draw[black!30] (1.25,-1.5) rectangle +(0.25,-0.25);
	\draw[black!30] (1.5,-1.5) rectangle +(0.25,-0.25);
	\draw[black!30] (1.75,-1.5) rectangle +(0.25,-0.25);
	\draw[black!30] (2,-1.5) rectangle +(0.25,-0.25);
	
	\draw[black!30][fill=black!10] (2.5,-0.25) rectangle +(2.25,0.5) node[color=black!80,midway] {\footnotesize Core 1};
	\draw[black!30][fill=black!10] (2.5,-0.5) rectangle +(1,-0.5) node[color=black!80,midway] {\footnotesize L1I};
	\draw[black!30] (3,-0.25) -- +(0,-0.25);
	\draw[black!30] (2.5,-1) rectangle +(0.25,-0.25);
	\draw[black!30] (2.75,-1) rectangle +(0.25,-0.25);
	\draw[black!30] (3,-1) rectangle +(0.25,-0.25);
	\draw[black!30] (3.25,-1) rectangle +(0.25,-0.25);
	\draw[black!30] (2.5,-1.25) rectangle +(0.25,-0.25);
	\draw[black!30] (2.75,-1.25) rectangle +(0.25,-0.25);
	\draw[black!30] (3,-1.25) rectangle +(0.25,-0.25);
	\draw[black!30] (3.25,-1.25) rectangle +(0.25,-0.25);
	\draw[black!30] (2.5,-1.5) rectangle +(0.25,-0.25);
	\draw[black!30] (2.75,-1.5) rectangle +(0.25,-0.25);
	\draw[black!30] (3,-1.5) rectangle +(0.25,-0.25);
	\draw[black!30] (3.25,-1.5) rectangle +(0.25,-0.25);
	\draw[black!30] (4.25,-0.25) -- +(0,-0.25);
	\draw[black!30,fill=black!10] (3.75,-0.5) rectangle +(1,-0.5) node[color=black!80,midway] {\footnotesize L1D};
	\draw[black!30] (3.75,-1) rectangle +(0.25,-0.25);
	\draw[black!30] (4,-1) rectangle +(0.25,-0.25);
	\draw[black!30] (4.25,-1) rectangle +(0.25,-0.25);
	\draw[black!30] (4.5,-1) rectangle +(0.25,-0.25);
	\draw[black!30,fill=green!90!black!50] (3.75,-1.25) rectangle +(0.25,-0.25);
	\draw[black!30,fill=green!70!black!50] (4,-1.25) rectangle +(0.25,-0.25);
	\draw[black!30,fill=green!95!black!50] (4.25,-1.25) rectangle +(0.25,-0.25);
	\draw[black!30,fill=green!75!black!50] (4.5,-1.25) rectangle +(0.25,-0.25);
	\draw[black!30] (3.75,-1.5) rectangle +(0.25,-0.25);
	\draw[black!30] (4,-1.5) rectangle +(0.25,-0.25);
	\draw[black!30] (4.25,-1.5) rectangle +(0.25,-0.25);
	\draw[black!30] (4.5,-1.5) rectangle +(0.25,-0.25);	
	
	\draw[black!30] (0.5,-1.75) -- +(0,-0.25);
	\draw[black!30] (1.75,-1.75) -- +(0,-0.25);
	\draw[black!30] (3,-1.75) -- +(0,-0.25);
	\draw[black!30] (4.25,-1.75) -- +(0,-0.25);
	\draw[black!30,fill=black!10] (0.375,-2) rectangle +(4,-0.5) node[color=black!80,midway] {\footnotesize L2 Unified Cache};
	\draw[black!30] [decorate,decoration={brace,amplitude=5pt}] (0.375,-3.25) -- (0.375,-2.5) node[color=black!80,yshift=+5pt,midway,sloped,above] {\footnotesize Sets};
	
	\draw [decorate,thick,decoration={brace,amplitude=5pt,aspect=0.3}] (4.75,-1.5) -- +(-1,0);
	
	\draw[black!30] (0.25+0.125,-2.5) rectangle +(0.25,-0.25);
	\draw[black!30] (0.5+0.125,-2.5) rectangle +(0.25,-0.25);
	\draw[black!30] (0.75+0.125,-2.5) rectangle +(0.25,-0.25);
	
	\draw[black!30] (1+0.125,-2.5) rectangle +(0.25,-0.25);
	\draw[black!30] (1.25+0.125,-2.5) rectangle +(0.25,-0.25);
	\draw[black!30] (1.5+0.125,-2.5) rectangle +(0.25,-0.25);
	\draw[black!30] (1.75+0.125,-2.5) rectangle +(0.25,-0.25);
	
	\draw[black!30] (2+0.125,-2.5) rectangle +(0.25,-0.25);
	\draw[black!30] (2.25+0.125,-2.5) rectangle +(0.25,-0.25);
	\draw[black!30] (2.5+0.125,-2.5) rectangle +(0.25,-0.25);
	\draw[black!30] (2.75+0.125,-2.5) rectangle +(0.25,-0.25);
	
	\draw[black!30] (3+0.125,-2.5) rectangle +(0.25,-0.25);
	\draw[black!30] (3.25+0.125,-2.5) rectangle +(0.25,-0.25);
	\draw[black!30] (3.5+0.125,-2.5) rectangle +(0.25,-0.25);
	\draw[black!30] (3.75+0.125,-2.5) rectangle +(0.25,-0.25);

	\draw[black!30] (4+0.125,-2.5) rectangle +(0.25,-0.25);
	
	\draw[black!30] (0.25+0.125,-2.75) rectangle +(0.25,-0.25);
	\draw[black!30] (0.5+0.125,-2.75) rectangle +(0.25,-0.25);
	\draw[black!30] (0.75+0.125,-2.75) rectangle +(0.25,-0.25);
	\draw[black!30,fill=red] (1+0.125,-2.75) rectangle +(0.25,-0.25);
	\draw[black!30] (1.25+0.125,-2.75) rectangle +(0.25,-0.25);
	\draw[black!30] (1.5+0.125,-2.75) rectangle +(0.25,-0.25);
	\draw[black!30] (1.75+0.125,-2.75) rectangle +(0.25,-0.25);
	\draw[black!30] (2+0.125,-2.75) rectangle +(0.25,-0.25);
	\draw[black!30] (2.25+0.125,-2.75) rectangle +(0.25,-0.25);
	\draw[black!30] (2.5+0.125,-2.75) rectangle +(0.25,-0.25);
	\draw[black!30] (2.75+0.125,-2.75) rectangle +(0.25,-0.25);
	\draw[black!30] (3+0.125,-2.75) rectangle +(0.25,-0.25);
	\draw[black!30,fill=green!80!black!50] (3.25+0.125,-2.75) rectangle +(0.25,-0.25);
	\draw[black!30,fill=green!60!black!50] (3.5+0.125,-2.75) rectangle +(0.25,-0.25);
	\draw[black!30,fill=green!65!black!50] (3.75+0.125,-2.75) rectangle +(0.25,-0.25);
	\draw[black!30,fill=green!85!black!50] (4+0.125,-2.75) rectangle +(0.25,-0.25);
	
	\draw[black!30] (0.25+0.125,-3) rectangle +(0.25,-0.25);
	\draw[black!30] (0.5+0.125,-3) rectangle +(0.25,-0.25);
	\draw[black!30] (0.75+0.125,-3) rectangle +(0.25,-0.25);
	\draw[black!30] (1+0.125,-3) rectangle +(0.25,-0.25);
	\draw[black!30] (1.25+0.125,-3) rectangle +(0.25,-0.25);
	\draw[black!30] (1.5+0.125,-3) rectangle +(0.25,-0.25);
	\draw[black!30] (1.75+0.125,-3) rectangle +(0.25,-0.25);
	\draw[black!30] (2+0.125,-3) rectangle +(0.25,-0.25);
	\draw[black!30] (2.25+0.125,-3) rectangle +(0.25,-0.25);
	\draw[black!30] (2.5+0.125,-3) rectangle +(0.25,-0.25);
	\draw[black!30] (2.75+0.125,-3) rectangle +(0.25,-0.25);
	\draw[black!30] (3+0.125,-3) rectangle +(0.25,-0.25);
	\draw[black!30] (3.25+0.125,-3) rectangle +(0.25,-0.25);
	\draw[black!30] (3.5+0.125,-3) rectangle +(0.25,-0.25);
	\draw[black!30] (3.75+0.125,-3) rectangle +(0.25,-0.25);
	\draw[black!30] (4+0.125,-3) rectangle +(0.25,-0.25);
	\draw[-latex',thick,out=180,in=270] (1+0.25,-2.75-0.125) to node[xshift=-0.075cm,yshift=0.075cm,midway,sloped,below]{\footnotesize (1)} (0.75-0.125,-1.25-0.125);
	\draw[-latex',thick,out=270,in=0] (4.45,-1.65) to node[yshift=-0.075cm,midway,sloped,above]{\footnotesize (2)} (4.25+0.125,-2.75-0.125);
	\draw[-latex',densely dashed,thick,out=180,in=0] (4.25,-2.75-0.125) to node[xshift=0.10cm,midway,sloped,below]{\footnotesize (3)} (1.5,-2.75-0.125);
	
	\end{tikzpicture}
	}
  \caption{Cross-core instruction cache eviction through data accesses.}
  \label{fig:data_to_instruction_eviction}
\end{figure}
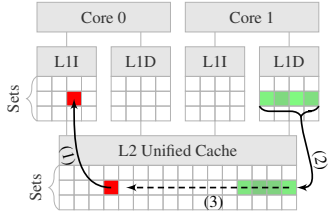

To perform a cross-core attack we load enough data into the cache to fully evict the corresponding last-level cache set. Thereby, we exploit that we can fill the last-level cache directly or indirectly depending on the cache organization. On the \Alcatel, the last-level cache is instruction-inclusive and thus we can evict instructions from the local caches of the other core. Figure~\ref{fig:data_to_instruction_eviction} illustrates such an eviction. In step 1, an instruction is allocated to the last-level cache and the instruction cache of one core. In step 2, a process fills its core's data cache, thereby evicting cache lines into the last-level cache. In step 3, the process has filled the last-level cache set using only data accesses and thereby evicts the instructions from instruction caches of other cores as well. 

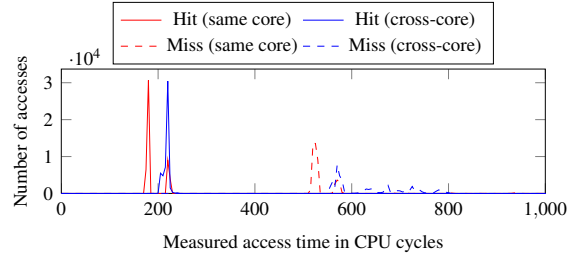
\begin{figure}[t]
\centering
\begin{tikzpicture}[scale=0.9]
\pgfplotsset{every axis legend/.append style={at={(0.5,1.54)},anchor=north}}
\begin{axis}[
legend columns=2, transpose legend,
style={font=\footnotesize},
xlabel=Measured access time in CPU cycles,
ylabel=Number of accesses,
width=1.1\hsize,
ymin=10,
xmin=0,
xmax=1000,
height=3.42cm,
unbounded coords=discard
]
\addplot+[no marks,red] table[x=Time,y=Hit] {hit_vs_miss_bacon.csv};
\addlegendentry{Hit (same core)}
\addplot+[no marks,red,dashed] table[x=Time,y=Miss] {hit_vs_miss_bacon.csv};
\addlegendentry{Miss (same core)}
\addplot+[no marks,blue] table[x=Time,y=Hit] {hit_vs_miss_bacon_cc.csv};
\addlegendentry{Hit (cross-core)}
\addplot+[no marks,blue,dashed] table[x=Time,y=Miss] {hit_vs_miss_bacon_cc.csv};
\addlegendentry{Miss (cross-core)}

\end{axis}
\end{tikzpicture}
 \caption{Histograms of cache hits and cache misses measured same-core and cross-core on the \OnePlus.}
  \label{fig:oneplus_hit_miss_cross_core}
\end{figure}

We access cache lines multiple times to perform transfers between L1 and L2 cache. Thus, more and more addresses used for eviction are cached in either L1 or L2. As ARM CPUs typically have L1 caches with a very low associativity, the probability of eviction to L2 through other system activity is high. Using an eviction strategy that performs frequent transfers between L1 and L2 increases this probability further. Thus, this approach also works for other cache organizations to perform cross-core and cross-CPU cache attacks. Due to the cache coherence protocol between the CPU cores~\cite{blog_a53,a53_manual}, remote-core fetches are faster than memory accesses and thus can be distinguished from cache misses. For instance, Figure~\ref{fig:oneplus_hit_miss_cross_core} shows the cache hit and miss histogram on the \OnePlus. 
The cross-core access introduces a latency of 40 CPU cycles on average. However, cache misses take more than 500 CPU cycles on average. Thus, cache hits and misses are clearly distinguishable based on a single threshold value.

\subsection{Fast Cache Eviction}\label{sec:eviction}
In this section, we tackle the aforementioned challenges 3 and 4, \ie not all ARM processors support a flush instruction, and the replacement policy is pseudo-random. 

There are two options to evict cache lines: (1) the flush instruction or (2) evict data with memory accesses to congruent addresses, \ie addresses that map to the same cache set. 
As the flush instruction is only available on the \Samsung, we need to rely on eviction strategies for the other devices and, therefore, to defeat the replacement policy. 
The L1 cache in Cortex-A53 and Cortex-A57 has a very small number of ways and employs a least-recently used (LRU) replacement policy~\cite{a57_manual}. However, for a full cache eviction, we also have to evict cache lines from the L2 cache, which uses a 
pseudo-random replacement policy.  

\paragraph{Eviction strategies.}
Previous approaches to evict data on Intel x86 platforms either have too much overhead~\cite{Hund2013} or are only applicable to caches implementing an LRU replacement policy~\cite{DBLP:conf/sp/LiuYGHL15,DBLP:conf/dimva/MauriceNHF15,DBLP:conf/ccs/OrenKSK15}. Spreitzer and Plos~\cite{DBLP:conf/cosade/SpreitzerP13} proposed an eviction strategy for ARMv7-A CPUs that requires to access more addresses than there are cache lines per cache set, due to the pseudo-random replacement policy. Recently, Gruss~\etal\cite{rowhammerjs} 
demonstrated how to automatically find fast eviction strategies on Intel x86 architectures. 
We show that their algorithm is applicable to ARM CPUs as well. Thereby, we establish eviction strategies in an automated way and significantly reduce the overhead compared to \cite{DBLP:conf/cosade/SpreitzerP13}. We evaluated more than 4\,200 access patterns on our smartphones and identified the best eviction strategies. Even though the cache employs a random replacement policy, average eviction rate and average execution time are reproducible. Eviction sets are computed based on physical addresses, which can be retrieved via \verb|/proc/self/pagemap| as current Android versions allow access to these mappings to any unprivileged app without any permissions. Thus, eviction patterns and eviction sets can be efficiently computed.

We applied the algorithm of Gruss~\etal\cite{rowhammerjs} to a set of physically congruent addresses.
Table~\ref{table:test_strategies} summarizes different eviction strategies, \ie loop parameters, for the Krait 400. $N$ denotes the total eviction set size (length of the loop), $A$ denotes the shift offset (loop increment) to be applied after each round, and $D$ denotes the number of memory accesses in each iteration (loop body).
The column \emph{cycles} states the average execution time in CPU cycles over 1 million evictions and the last column denotes the average eviction rate.
The first line in Table~\ref{table:test_strategies} shows the average execution time and the average eviction rate for the privileged flush instruction, which gives the best result in terms of average execution time ($549$ CPU cycles). We evaluated $1\,863$ different strategies and our best identified eviction strategy ($N=11$, $A=2$, $D=2$) also achieves an average eviction rate of $100\%$ but takes $1\,578$ CPU cycles.
Although a strategy accessing every address in the eviction set only once ($A=1$, $D=1$, also called LRU eviction) performs significantly fewer memory accesses, it consumes more CPU cycles. For an average eviction rate of $100\%$, LRU eviction requires an eviction set size of at least $16$. The average execution time then is $3\,026$ CPU cycles.
Considering the eviction strategy used in~\cite{DBLP:conf/cosade/SpreitzerP13} that takes $4\,371$ CPU cycles, clearly demonstrates the advantage of our optimized eviction strategy that takes only $1\,578$ CPU cycles. 

\setlength{\tabcolsep}{6pt}
\begin{table}
  \centering
  \caption{Different eviction strategies on the Krait 400.}
  \vspace{5pt}
\resizebox{0.7\columnwidth}{!}{
\small{
  \begin{tabular}{rrrrr}
  \toprule 
\, \, $N$ & $A$ & $D$ & \, \, \, Cycles & Eviction rate \\
  \midrule
- & - & - & 549 & 100.00\% \\
11 & 2 & 2 & 1\,578 & 100.00\% \\
12 & 1 & 3 & 2\,094 & 100.00\% \\
13 & 1 & 5 & 2\,213 & 100.00\% \\
16 & 1 & 1 & 3\,026 & 100.00\% \\
24 & 1 & 1 & 4\,371 & 100.00\% \\
13 & 1 & 2 & 2\,372 & 99.58\% \\
11 & 1 & 3 & 1\,608 & 80.94\% \\
11 & 4 & 1 & 1\,948 & 58.93\% \\
10 & 2 & 2 & 1\,275 & 51.12\% \\
  \bottomrule
 \end{tabular}
 }
}
  \label{table:test_strategies}
\end{table}

We performed the same evaluation with $2\,295$ different strategies on the ARM Cortex-A53 in our \Alcatel test system and summarize them in Table~\ref{table:test_strategies_a53}. For the best strategy we found ($N=21$, $A=1$, $D=6$), we measured an average eviction rate of $99.93\%$ and an average execution time of $4\,275$ CPU cycles.
We observed that LRU eviction ($A=1$, $D=1$) on the ARM Cortex-A53 would take 28 times more CPU cycles to achieve an average eviction rate of only $99.10\%$, thus it is not suitable for attacks on the last-level cache as used in previous work~\cite{DBLP:conf/cosade/SpreitzerP13}. The reason for this is that data can only be allocated to L2 cache by evicting it from the L1 cache on the ARM Cortex-A53. Therefore, it is better to reaccess the data that is already in the L2 cache and gradually add new addresses to the set of cached addresses instead of accessing more different addresses.

\begin{table}
  \centering
  \caption{Different eviction strategies on the Cortex-A53.}
  \vspace{5pt}
\resizebox{.7\columnwidth}{!}{
\small{
  \begin{tabular}{rrrrr}
  \toprule 
\, \, $N$ & $A$ & $D$ & \, \, \, Cycles & Eviction rate \\
  \midrule
- & - & - & 767 & 100.00\% \\
23 & 2 & 5 & 6\,209 & 100.00\% \\
23 & 4 & 6 & 16\,912 & 100.00\% \\
22 & 1 & 6 & 5\,101 & 99.99\% \\
21 & 1 & 6 & 4\,275 & 99.93\% \\
20 & 4 & 6 & 13\,265 & 99.44\% \\
800 & 1 & 1 & 142\,876 & 99.10\% \\
200 & 1 & 1 & 33\,110 & 96.04\% \\
100 & 1 & 1 & 15\,493 & 89.77\% \\
48 & 1 & 1 & 6\,517 & 70.78\% \\
  \bottomrule
 \end{tabular}
 }
}
  \label{table:test_strategies_a53}
\end{table}

\begin{figure}[t]
\centering
\begin{tikzpicture}[scale=0.9]
\pgfplotsset{every axis legend/.append style={at={(0.5,1.54)},anchor=north}}
\begin{axis}[
legend columns=2, transpose legend,
style={font=\footnotesize},
xlabel=Measured execution time in CPU cycles,
ylabel=Number of cases,
width=1.1\hsize,
ymin=10,
xmin=0,
xmax=600,
height=3.42cm,
unbounded coords=discard
]
\addplot+[no marks,blue] table[x=Time,y=Hit] {hit_vs_miss_s6_ff_perf.csv};
\addlegendentry{Flush (address cached)}
\addplot+[no marks,red,dashed] table[x=Time,y=Miss] {hit_vs_miss_s6_ff_perf.csv};
\addlegendentry{Flush (address not cached)}

\end{axis}
\end{tikzpicture}
 \caption{Histograms of the execution time of the flush operation on cached and
 not cached addresses measured on the \Samsung.}
  \label{fig:s6_hit_miss_ff}
\end{figure}
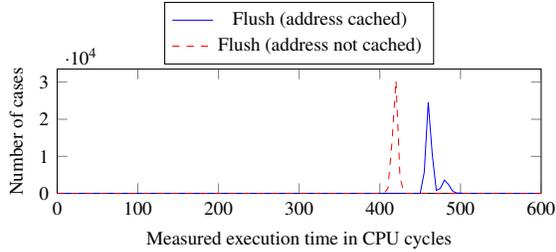

On the ARM Cortex-A57 the userspace flush instruction was significantly faster in any case. Thus, for \FlushReload we use the flush instruction and for \PrimeProbe the eviction strategy. Falling back to \EvictReload is not necessary on the Cortex-A57. Similarly to recent Intel x86 CPUs, the execution time of the flush instruction on ARM depends on whether or not the value is cached, as shown in Figure~\ref{fig:s6_hit_miss_ff}. 
The execution time is higher if the address is cached and lower if the address is not cached. This observation allows us to distinguish between cache hits and cache misses depending on the timing behavior of the flush instruction, and therefore to perform a \FlushOnly attack. Thus, in case of shared memory between the victim and the attacker, it is not even required to evict and reload an address in order to exploit the cache side channel.

\paragraph{A note on \PrimeProbe.}
Finding a fast eviction strategy for \PrimeProbe on architectures with a random replacement policy is not as straightforward 
as on Intel x86.
Even in case of x86 platforms, the problem of cache trashing has been discussed by Tromer~\etal\cite{DBLP:journals/joc/TromerOS10}. Cache trashing occurs when reloading (probing) an address evicts one of the addresses that are to be accessed next. While Tromer~\etal were able to overcome this problem by using a doubly-linked list that is accessed forward during the prime step and backwards during the probe step, the random replacement policy on ARM also contributes to the negative effect of cache trashing. 

We analyzed the behavior of the cache and designed a prime step and a probe step that work with a smaller set size to avoid set thrashing. Thus, we set the eviction set size to 15 on the \Alcatel. As we run the \PrimeProbe attack in a loop, exactly 1 way in the L2 cache will not be occupied after a few attack rounds. We might miss a victim access in $\frac{1}{16}$ of the cases, which however is necessary as otherwise we would not be able to get reproducible measurements at all due to set thrashing. If the victim replaces one of the 15 ways occupied by the attacker, there is still one free way to reload the address that was evicted. This reduces the chance of set thrashing significantly and allows us to successfully perform \PrimeProbe on caches with a random replacement policy.

\subsection{Accurate Unprivileged Timing}\label{sec:cyclecount}
In this section, we tackle the aforementioned challenge 5, \ie cycle-accurate timings require root access on ARM. 

In order to distinguish cache hits and cache misses, timing sources or dedicated performance counters can be used. 
We focus on timing sources, as cache misses have a significantly higher access latency and timing sources are well studied on Intel x86 CPUs.
Cache attacks on x86 CPUs employ the unprivileged \texttt{rdtsc} instruction to obtain a sub-nanosecond resolution timestamp. 
The ARMv7-A architecture does not provide an instruction for this purpose. Instead, the ARMv7-A architecture has a performance monitoring unit that allows to monitor CPU activity. One of these performance counters---denoted as \emph{cycle count register} (PMCCNTR)---can be used to distinguish cache hits and cache misses by relying on the number of CPU cycles that passed during a memory access. However, these performance counters are not accessible from userspace by default and an attacker would need root privileges. 

We broaden the attack surface by exploiting timing sources that are accessible without any privileges or permissions. We identified three possible alternatives for timing measurements. 
\begin{description}[leftmargin=10pt]
 \item[Unprivileged syscall.] The \verb|perf_event_open| syscall is an
abstract layer to access performance information through the kernel independently of the underlying
hardware. For instance, \verb|PERF_COUNT_HW_CPU_CYCLES| returns an accurate cycle count including a minor overhead due to the syscall.
The availability of this feature depends on the Android kernel configuration, \eg the stock kernel on the \Alcatel as well as the \OnePlus provide this feature by default.
Thus, in contrast to previous work~\cite{DBLP:conf/cosade/SpreitzerP13}, the attacker does not have to load a kernel module to access this information as the \verb|perf_event_open| syscall can be accessed without any privileges or permissions.
 
 \item[POSIX function.] Another alternative to obtain sufficiently accurate timing information is the POSIX function \verb|clock_gettime()|, 
 with an accuracy in the range of microseconds to nanoseconds. Similar information can also be obtained from \verb|/proc/timer_list|.

 \item[Dedicated thread timer.] If no interface with sufficient accuracy is available, an attacker can run a thread that increments a global variable in a loop, providing a fair approximation of a cycle counter. Our experiments show that this approach works reliably on smartphones as well as recent x86 CPUs. The resolution of this threaded timing information is as high as with the other methods.

\end{description}

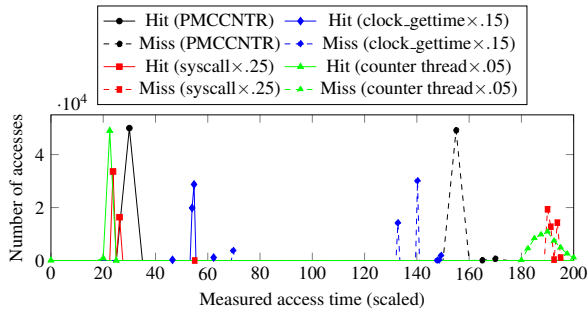
\begin{figure}[t]
\centering
\begin{tikzpicture}[scale=0.7]
\pgfplotsset{every axis legend/.append style={at={(0.5,1.75)},anchor=north}}
\begin{axis}[
legend columns=4, transpose legend, 
xlabel=Measured access time (scaled),
ylabel=Number of accesses,
width=1.45\hsize,
ymin=10,
xmin=0,
xmax=200,
height=4.35cm,
unbounded coords=discard
]
\addplot+[mark size=1.5pt,mark=*,solid,black,draw=black,mark options={draw=black,fill=black}] table[x expr=\thisrow{Time},y=Hit] {hit_vs_miss_alto45_reg.csv};
\addlegendentry{Hit (PMCCNTR)}
\addplot+[mark size=1.5pt,mark=*,dashed,black,draw=black,mark options={draw=black,fill=black}] table[x expr=\thisrow{Time},y=Miss] {hit_vs_miss_alto45_reg.csv};
\addlegendentry{Miss (PMCCNTR)}

\addplot+[mark size=1.5pt,mark=square*,solid,red,draw=red,mark options={draw=red,fill=red}] table[x expr=\thisrow{Time}*0.25,y=Hit] {hit_vs_miss_alto45_perf.csv};
\addlegendentry{Hit (syscall$\times.25$)}
\addplot+[mark size=1.5pt,mark=square*,dashed,red,draw=red,mark options={draw=red,fill=red}] table[x expr=\thisrow{Time}/4,y=Miss] {hit_vs_miss_alto45_perf.csv};
\addlegendentry{Miss (syscall$\times.25$)}

\addplot+[mark size=2pt,mark=diamond*,solid,blue,draw=blue,mark options={draw=blue,fill=blue}] table[x expr=\thisrow{Time}*0.15,y=Hit] {hit_vs_miss_alto45_monotonic.csv};
\addlegendentry{Hit (clock\_gettime$\times.15$)}
\addplot+[mark size=2pt,mark=diamond*,dashed,blue,draw=blue,mark options={draw=blue,fill=blue}] table[x expr=\thisrow{Time}*0.15,y=Miss] {hit_vs_miss_alto45_monotonic.csv};
\addlegendentry{Miss (clock\_gettime$\times.15$)}

\addplot+[mark size=1.8pt,mark=triangle*,solid,green,draw=green,mark options={draw=green,fill=green}] table[x expr=\thisrow{Time}*0.05,y=Hit] {hit_vs_miss_alto45_threadcounter.csv};
\addlegendentry{Hit (counter thread$\times.05$)}
\addplot+[mark size=1.8pt,dashed,mark=triangle*,green,draw=green,mark options={draw=green,fill=green}] table[x expr=\thisrow{Time}*0.05,y=Miss] {hit_vs_miss_alto45_threadcounter.csv};
\addlegendentry{Miss (counter thread$\times.05$)}

\end{axis}
\end{tikzpicture}
 \caption{Histogram of cross-core cache hits/misses on the \Alcatel using different methods. X-values are scaled for visual representation.}
  \label{fig:kernel_module_perf_difference}
\end{figure}

In Figure~\ref{fig:kernel_module_perf_difference} we show the cache hit and miss histogram based on the four different methods, including the cycle count register, on a \Alcatel. Despite the latency and noise, cache hits and cache misses are clearly distinguishable with all approaches. Thus, all methods can be used to implement cache attacks. Determining the best timing method on the device under attack can be done in a few seconds during an online attack. 

\section{High Performance Covert Channels}\label{sec:covert} 
To evaluate the performance of our attacks, we measure the capacity of cross-core and cross-CPU cache covert channels. A covert channel enables two unprivileged applications on a system to communicate with each other without using any data transfer mechanisms provided by the operating system. This communication evades the sandboxing concept and the permission system (\cf collusion attacks~\cite{DBLP:conf/acsac/MarforioRFC12}). Both applications were running in the background while the phone was mostly idle and an unrelated app was running as the foreground application.

\begin{table*}[t]
\caption{Comparison of covert channels on Android.}
  \vspace{5pt}
\centering
\resizebox{.7\hsize}{!}{
\small{
  \begin{tabular}{llrr}
  \toprule 
  Work                                                     & Type                              & Bandwidth [bps]    & Error rate\\
  \midrule
  Ours (\Samsung)                                          & \FlushReload, cross-core   & \textbf{1\,140\,650} & 1.10\% \\
  Ours (\Samsung)                                          & \FlushReload, cross-CPU    & \textbf{257\,509} & 1.83\% \\
  Ours (\Samsung)                                          & \FlushOnly,   cross-core   & \textbf{178\,292} & 0.48\% \\
  Ours (\Alcatel)                                          & \EvictReload, cross-core   & \textbf{13\,618} & 3.79\%\\
  Ours (\OnePlus)                                          & \EvictReload, cross-core   & \textbf{12\,537} & 5.00\% \\
  Marforio~\etal\cite{DBLP:conf/acsac/MarforioRFC12}       & Type of Intents                   & 4\,300 & -- \\ 
  Marforio~\etal\cite{DBLP:conf/acsac/MarforioRFC12}       & UNIX socket discovery             & 2\,600 & --\\ 
  Schlegel~\etal\cite{DBLP:conf/ndss/SchlegelZZIKW11}      & File locks                        & 685 & -- \\
  Schlegel~\etal\cite{DBLP:conf/ndss/SchlegelZZIKW11}      & Volume settings                   & 150 & -- \\
  Schlegel~\etal\cite{DBLP:conf/ndss/SchlegelZZIKW11}      & Vibration settings                & 87 & --\\
  \bottomrule
 \end{tabular}
 }
}
\label{tab:cov_channels}
\end{table*}

Our covert channel is established on addresses of a shared library that is used by both the sender and the receiver. While both processes have read-only access to the shared library,
they can transmit information by loading addresses from the shared library into the cache or evicting (flushing) it from the cache, respectively.

The covert channel transmits packets of $n$-bit data, an $s$-bit sequence number, and a $c$-bit checksum that is
computed over data and sequence number. The sequence number is used
to distinguish consecutive packets and the checksum is used to check the
integrity of the packet. The receiver acknowledges valid packets by responding with an $s$-bit sequence number and an $x$-bit checksum. By adjusting the sizes of checksums and sequence numbers the error rate of the covert channel can be controlled.

Each bit is represented by one address in the shared library, whereas no two addresses are chosen that map to the same cache set. To transmit a bit value of 1, the sender accesses the corresponding address in the library. To transmit a bit value of 0, the sender does not access the corresponding address, resulting in a cache miss on the receiver's side. Thus, the receiving process observes a cache hit or a cache miss depending on the memory access performed by the sender. The same method is used for the acknowledgements sent by the receiving process.

We implemented this covert channel using \EvictReload, \FlushReload, and
\FlushOnly on our smartphones. The results are summarized in Table~\ref{tab:cov_channels}. 
On the \Samsung, we achieve a cross-core transmission rate of 1\,140\,650\,bps at an error rate of $1.10\%$. This is 265 times faster than any existing covert channel on smartphones. In a cross-CPU transmission we achieve a transmission rate of 257\,509\,bps at an error rate of $1.83\%$. We achieve a cross-core transition
rate of 178\,292\,bps at an error rate of $0.48\%$ using \FlushOnly on the
\Samsung.
On the \Alcatel we achieve a cross-core transmission rate of 13\,618\,bps at an
error rate of $3.79\%$ using \EvictReload. This is still 3 times faster than previous covert channels on
smartphones. The covert channel is significantly slower on the \Alcatel than on the \Samsung because the hardware is much slower, \EvictReload is slower than \FlushReload, and retransmission might be necessary in $0.14\%$ of the cases where eviction is not successful (cf. Section~\ref{sec:eviction}). On the older \OnePlus we achieve a cross-core
transmission rate of 12\,537\,bps at an error rate of $5.00\%$, 3 times faster than
previous covert channels on smartphones.
The reason for the higher error rate is the additional timing noise due to the cache coherence protocol performing a high number of remote-core fetches.

\section{Attacking User Input on Smartphones}\label{sec:userinput}
In this section we demonstrate cache side-channel attacks on Android smartphones. We implement cache template attacks~\cite{DBLP:conf/uss/GrussSM15} to create and exploit accurate cache-usage profiles using the \EvictReload or \FlushReload attack.
Cache template attacks have a profiling phase and an exploitation phase. In the profiling phase, a template matrix is computed that
represents how many cache hits occur on a specific address when triggering a specific event.
The exploitation phase uses this matrix to infer events from cache hits.

To perform cache template attacks, an attacker has to map shared binaries or shared libraries as read-only shared memory into its own address space. By using shared libraries, the attacker bypasses any potential countermeasures taken by the operating system, such as restricted access to runtime data of other apps or address space layout randomization (ASLR). The attack can even be performed online on the device under attack if the event can be simulated.

Triggering the actual event that an attacker wants to spy on might require either (1) an offline phase or (2) privileged access. For instance, in case of a keylogger, the attacker can gather a cache template matrix offline for a specific version of a library, or the attacker relies on privileged access of the application (or a dedicated permission) in order to be able to simulate events for gathering the cache template matrix. However, the actual exploitation of the cache template matrix to infer events neither requires privileged access nor any permission.

\subsection{Attacking a Shared Library}
Just as Linux, Android uses a large number of shared libraries, each with a size of up to several megabytes.
We inspected all available libraries on the system by manually scanning the names and identified libraries that 
might be responsible for handling user input, \eg the \texttt{libinput.so} library. Without loss of generality, we restricted the set of attacked libraries since testing all libraries would have taken a significant amount of time. Yet, an adversary could exhaustively probe all libraries.

We automated the search for addresses in these shared libraries and after identifying addresses, we monitored them in order to infer user input events. 
For instance, in the profiling phase on \texttt{libinput.so}, we simulated events via the android-debug bridge (adb shell) with two different methods. The first method uses the \texttt{input} command line tool to simulate user input events. The second method is writing event messages to \verb|/dev/input/event*|. Both methods can run entirely on the device for instance in idle periods while the user is not actively using the device.
As the second method only requires a \verb|write()| statement it is significantly faster, but it is also more device specific. Therefore, we used the \texttt{input} command line except when profiling differences between different letter keys.
While simulating these events, 
we simultaneously probed all addresses within the \texttt{libinput.so} library, \ie we measured the number of cache hits that occurred on each address
when triggering a specific event. As already mentioned above, the simulation of some events might require either an offline phase or specific privileges in case of online attacks. 

\begin{figure}[t]
  \centering
  \includegraphics[width=0.84\hsize]{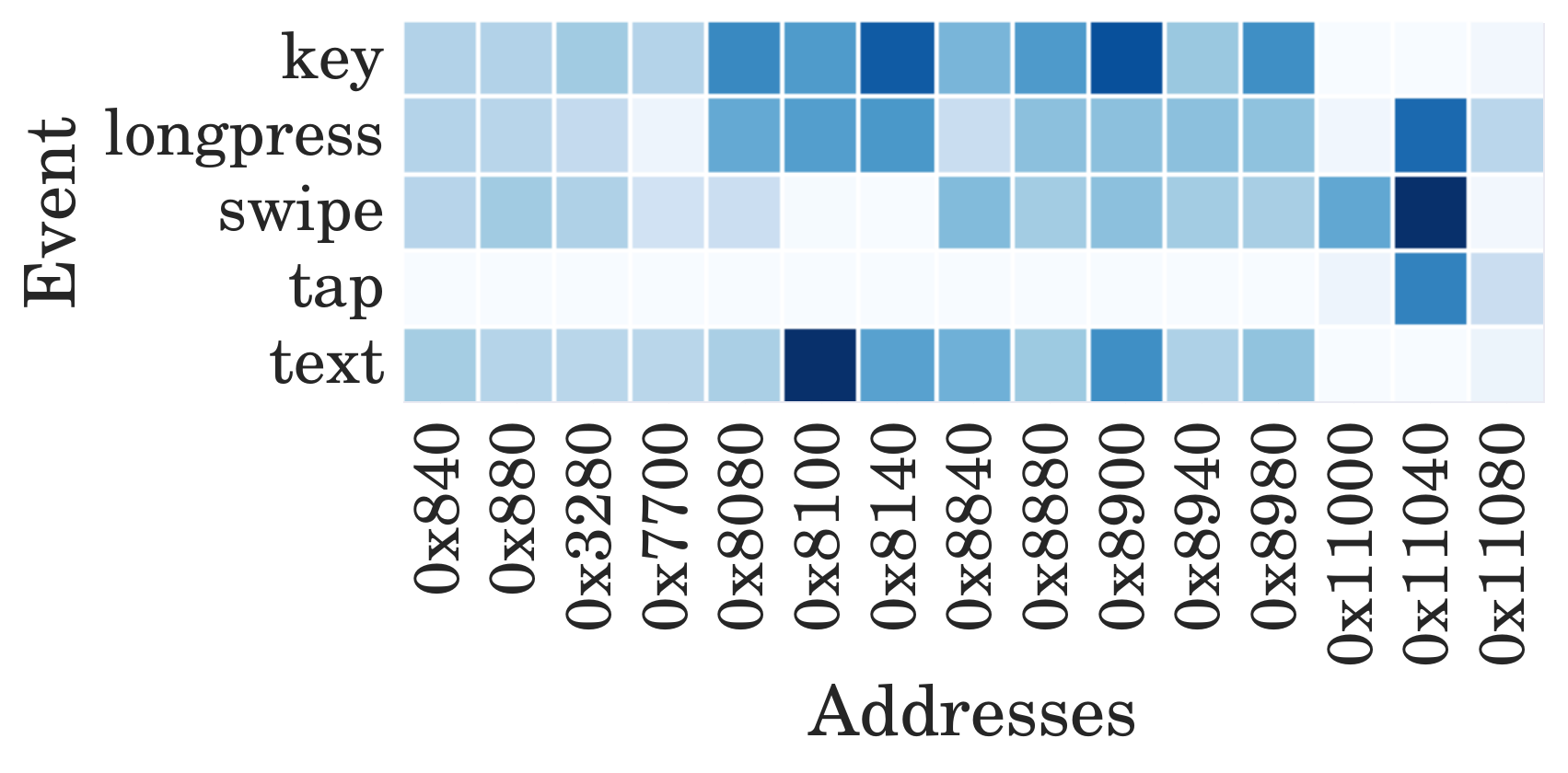}
  
  \caption{Cache template matrix for \texttt{libinput.so}.}
  \label{fig:ta_heatmap_libinput}
\end{figure}

Figure~\ref{fig:ta_heatmap_libinput} shows part of the cache template matrix for 
\texttt{libinput.so}. 
We triggered the following events: key events including the power button (\emph{key}), long touch events (\emph{longpress}), \emph{swipe} events, touch events (\emph{tap}), and text input events (\emph{text}) via 
the \texttt{input} 
tool as often as possible and measured each address and event for one second.
The cache template matrix clearly reveals addresses with high cache-hit rates for 
specific events. 
Darker colors represent addresses with higher cache-hit rates for a specific event and 
lighter colors represent addresses with lower cache-hit rates.
Hence, we can distinguish different events based on cache hits on these addresses.

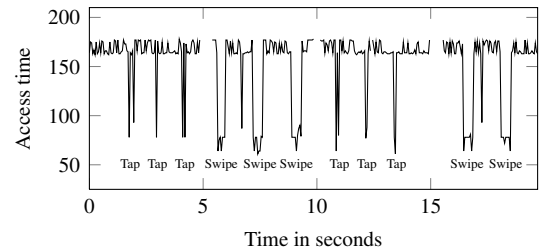
\begin{figure}[t]
\centering
\vspace{-2.2pt}
\begin{tikzpicture}
\begin{axis}[
style={font=\footnotesize},
xlabel=Time in seconds,
ylabel=Access time,
width=0.95\hsize,
xmin=0,
xmax=19.7000,
ymax=210,
ymin=25, 
height=4cm
]
\addplot[no marks,restrict x to domain=0:4.85] table[x=Time,y=Value] {tap-vs-swipe2.log};
\addplot[no marks,restrict x to domain=5.4:9.85] table[x=Time,y=Value] {tap-vs-swipe2.log};
\addplot[no marks,restrict x to domain=10.15:12.35] table[x=Time,y=Value] {tap-vs-swipe2.log};
\addplot[no marks,restrict x to domain=12.45:14.95] table[x=Time,y=Value] {tap-vs-swipe2.log};
\addplot[no marks,restrict x to domain=15.55:19.7] table[x=Time,y=Value] {tap-vs-swipe2.log};
\node [] at (axis cs:  1.8,  50) {\tiny{Tap}};
\node [] at (axis cs:  3,  50) {\tiny{Tap}};
\node [] at (axis cs:  4.2,  50) {\tiny{Tap}};

\node [] at (axis cs:  5.8,  50) {\tiny{Swipe}};
\node [] at (axis cs:  7.5,  50) {\tiny{Swipe}};
\node [] at (axis cs:  9.1,  50) {\tiny{Swipe}};

\node [] at (axis cs:  11,  50) {\tiny{Tap}};
\node [] at (axis cs:  12.2,  50) {\tiny{Tap}};
\node [] at (axis cs:  13.5,  50) {\tiny{Tap}};

\node [] at (axis cs:  16.6,  50) {\tiny{Swipe}};
\node [] at (axis cs:  18.3,  50) {\tiny{Swipe}};
\end{axis}
\end{tikzpicture}
  \caption{Monitoring address \texttt{0x11040} of \texttt{libinput.so} on the \Alcatel reveals taps and swipes.} 
  \label{fig:ta_swipe_vs_tap}
\end{figure}

We verified our results by monitoring the identified addresses while operating the smartphone manually, \ie we touched the screen and our attack application reliably reported cache hits on the monitored addresses. For instance, address \verb|0x11040| of \verb|libinput.so| can be used to distinguish tap actions and swipe actions on the screen of the \Alcatel. Tap actions cause a smaller number of cache hits than swipe actions. Swipe actions cause cache hits in a high frequency as long as the screen is touched.  
Figure~\ref{fig:ta_swipe_vs_tap} shows a sequence of 3 tap events, 3 swipe events, 3 tap events, and 2 swipe events. These events can be clearly distinguished due to the fast access times.
The gaps mark periods of time where our program was not scheduled on the CPU. Events occurring in those periods can be missed by our attack.

Swipe input allows 
to enter words by swiping over the soft-keyboard and thereby connecting single characters to form a word. Since we are able to determine the length of swipe movements, we can correlate the length of the swipe movement with the actual word length 
in any Android application or system interface that uses swipe input without any privileges. Furthermore, we can determine the actual length of the unlock pattern for the pattern-unlock mechanism.

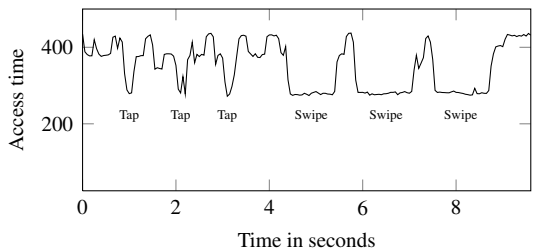
\begin{figure}[h!]
\centering
\begin{tikzpicture}
\begin{axis}[
style={font=\footnotesize},
xlabel=Time in seconds,
ylabel=Access time,
width=0.95\hsize,
xmin=0,
xmax=9.6,
ymax=500,
ymin=25, 
height=4cm
]
\addplot[no marks] table[x=Time,y=Value] {tap-vs-swipe-s6.csv};
\node [] at (axis cs:  1.0,  220) {\tiny{Tap}};
\node [] at (axis cs:  2.1,  220) {\tiny{Tap}};
\node [] at (axis cs:  3.1,  220) {\tiny{Tap}};
\node [] at (axis cs:  4.9,  220) {\tiny{Swipe}};
\node [] at (axis cs:  6.5,  220) {\tiny{Swipe}};
\node [] at (axis cs:  8.1,  220) {\tiny{Swipe}};
\end{axis}
\end{tikzpicture}
 \caption{Monitoring address \texttt{0xDC5C} of \texttt{libinput.so} on the \Samsung reveals tap and swipe events.}
  \label{fig:ta_swipe_vs_tap_s6}
\end{figure}

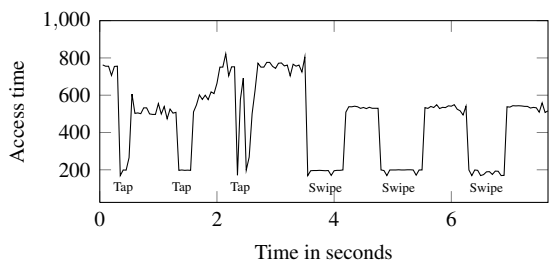
\begin{figure}[h!]
\centering
\begin{tikzpicture}
\begin{axis}[
style={font=\footnotesize},
xlabel=Time in seconds,
ylabel=Access time,
width=0.95\hsize,
xmin=0,
xmax=7.65,
ymax=1000,
ymin=25,
height=4cm
]
\addplot[no marks] table[x=Time,y=Value] {tap-vs-swipe-bacon.csv};
\node [] at (axis cs:  0.4,  100) {\tiny{Tap}};
\node [] at (axis cs:  1.4,  100) {\tiny{Tap}};
\node [] at (axis cs:  2.4,  100) {\tiny{Tap}};
\node [] at (axis cs:  3.85, 100) {\tiny{Swipe}};
\node [] at (axis cs:  5.1,  100) {\tiny{Swipe}};
\node [] at (axis cs:  6.6,  100) {\tiny{Swipe}};
\end{axis}
\end{tikzpicture} 
\caption{Monitoring address \texttt{0xBFF4} of \texttt{libinput.so} on the \OnePlus reveals tap and swipe events.}
  \label{fig:ta_swipe_vs_tap_bacon}
\end{figure}

Figure~\ref{fig:ta_swipe_vs_tap_s6} shows a user input sequence consisting of 3 tap events and 3 swipe events on the \Samsung. The attack was conducted using \FlushReload.
An attacker can monitor every single event. Taps and swipes can be distinguished based on the length of the cache hit phase. The length of a swipe movement can be determined from the same information.
Figure~\ref{fig:ta_swipe_vs_tap_bacon} shows the same experiment on the \OnePlus using \EvictReload. Thus, our attack techniques work on coherent non-inclusive last-level caches.

\subsection{Attacking ART Binaries}
\begin{figure}
  \centering
  \includegraphics[width=0.67\hsize]{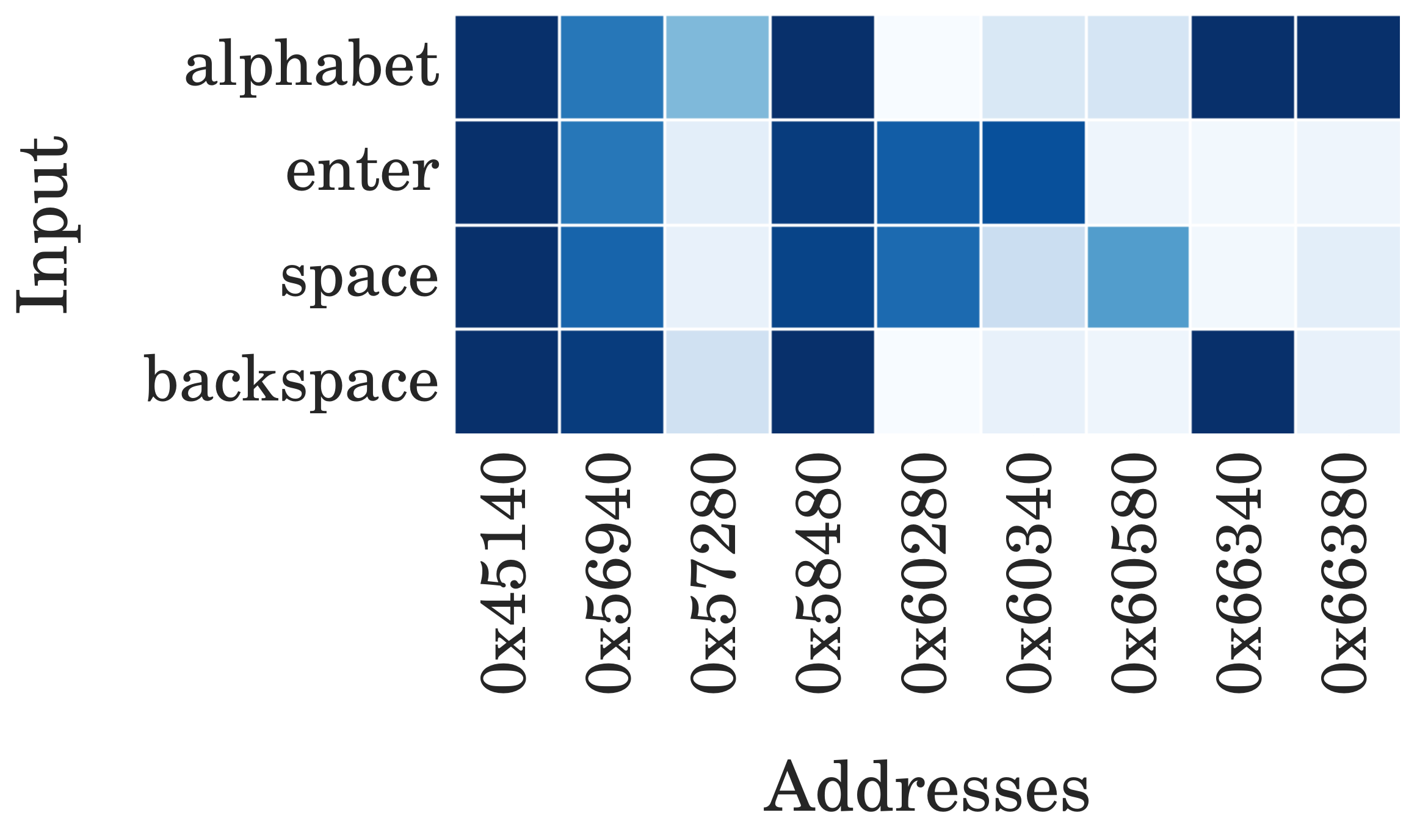} \caption{Cache template matrix for the default AOSP keyboard.} 
  \label{fig:ta_aosp_kb}
\end{figure}

Instead of attacking shared libraries, it is also possible to apply this attack to ART (Android Runtime) executables~\cite{art} that are compiled ahead of time. We used this attack on the
default AOSP keyboard and evaluated the number of accesses to every address in
the optimized executable that responds to an input of a letter on the
keyboard. It is possible to find addresses that correspond to a key press and
more importantly to distinguish between taps and key presses. Figure~\ref{fig:ta_aosp_kb}
shows the corresponding cache template matrix. We summarize the letter keys in one line (\emph{alphabet}) as they did not vary significantly. These addresses can be used to monitor
key presses on the keyboard. We identified an address that
corresponds only to letters on the keyboard and hardly on the space bar or the
return button. With this information it is possible to precisely determine the
length of single words entered using the default AOSP keyboard.

\begin{figure}[t]
\centering
\begin{tikzpicture}
\begin{axis}[
style={font=\footnotesize},
xlabel=Time in seconds,
ylabel=Access time,
width=0.95\hsize,
xtick={0,1,2,3,4,5,6,7,8},
xmin=0,
xmax=7.5,
ymax=360,
ymin=40,
height=4cm
]
\addplot+[blue,no marks] table[x=time,y=key] {keyboard_final.log};
\addlegendentry{Key}
\addplot+[red,only marks,mark=*,restrict y to domain=0:150,mark options={draw=red,fill=red,opacity=0.5},mark size=1pt] table[x=time,y=space] {keyboard_final.log};
\addlegendentry{Space}
\addplot+[red,only marks,mark=*,mark options={draw=red,fill=red,opacity=0.2},mark size=1pt] table[x=time,y=space] {keyboard_final.log};

\node [anchor=base] at (axis cs:  0.67,  80) {\tiny{t}};
\node [anchor=base] at (axis cs:  0.92,  80) {\tiny{h}};
\node [anchor=base] at (axis cs:  1.15,  80) {\tiny{i}};
\node [anchor=base] at (axis cs:  1.42,  80) {\tiny{s}};
\node [anchor=base] at (axis cs:  1.95,  80) {\tiny{\textit{Space}}};
\node [anchor=base] at (axis cs:  2.54,  80) {\tiny{i}};
\node [anchor=base] at (axis cs:  2.8,  80) {\tiny{s}};
\node [anchor=base] at (axis cs:  3.35,  80) {\tiny{\textit{Space}}};
\node [anchor=base] at (axis cs:  3.93,  80) {\tiny{a}};
\node [anchor=base] at (axis cs:  4.37,  80) {\tiny{\textit{Space}}};
\node [anchor=base] at (axis cs:  4.94,  80) {\tiny{m}};
\node [anchor=base] at (axis cs:  5.22,  80) {\tiny{e}};
\node [anchor=base] at (axis cs:  5.48,  80) {\tiny{s}};
\node [anchor=base] at (axis cs:  5.84,  80) {\tiny{s}};
\node [anchor=base] at (axis cs:  6.21,  80) {\tiny{a}};
\node [anchor=base] at (axis cs:  6.47,  80) {\tiny{g}};
\node [anchor=base] at (axis cs:  6.81,  80) {\tiny{e}};
\end{axis}
\end{tikzpicture}
  \caption{\EvictReload on 2 addresses in \texttt{custpack@\allowbreak app@\allowbreak withoutlibs@\allowbreak LatinIME.apk@\allowbreak classes.dex} on the \Alcatel while entering the sentence ``this is a message''.}
  \label{fig:ta_swipe_sentence}
\end{figure}
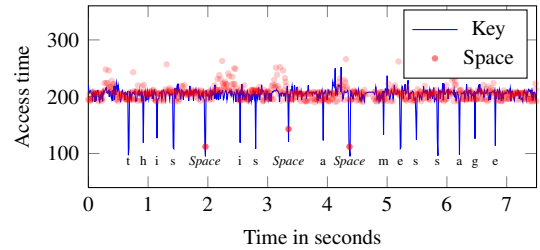

We illustrate the capability of detecting word lengths in Figure~\ref{fig:ta_swipe_sentence}. The blue line shows the timing measurements for the address identified for keys in general, the red dots represent measurements of the address for the space key. The plot shows that we can clearly determine the length of entered words and monitor user input accurately over time.

\subsection{Discussion and Impact}
Our proof-of-concept attacks exploit shared libraries and binaries from Android apk files to infer key strokes. The cache template attack technique we used for these attacks is generic and can also be used to attack any other library. For instance, there are various libraries that handle different hardware modules and software events on the device, such as GPS, Bluetooth, camera, NFC, vibrator, audio and video decoding, web and PDF viewers. Each of these libraries contains code that is executed and data that is accessed when the device is in use.
Thus, an attacker can perform a cache template attack on any of these libraries and spy on the corresponding device events. For instance, our attack can be used to monitor activity of the GPS sensor, bluetooth, or the camera. An attacker can record such user activities over time to learn more about the user.

We can establish inter-keystroke timings at an accuracy 
as high as the accuracy of cache side-channel attacks on keystrokes on x86 systems with a physical keyboard.
Thus, 
the inter-keystroke timings 
can be used to infer entered words, as has been shown by Zhang~\etal\cite{DBLP:conf/uss/ZhangW09}. Our attack even has a higher resolution than \cite{DBLP:conf/uss/ZhangW09}, \ie it is sub-microsecond accurate. 
Furthermore, we can distinguish between keystrokes on the soft-keyboard and generic touch actions outside the soft-keyboard. This information can 
be used to enhance sensor-based keyloggers that 
infer user input on mobile devices by exploiting, \eg 
the accelerometer and the gyroscope~\cite{DBLP:conf/uss/CaiC11,DBLP:conf/trust/CaiC12,DBLP:conf/acsac/AvivSBS12,DBLP:conf/mobisys/MiluzzoVBC12,DBLP:conf/wisec/XuBZ12} or the ambient-light sensor~\cite{DBLP:conf/ccs/Spreitzer14}. However, 
these attacks suffer from a lack of knowledge when exactly a user touches the screen. Based on our attack, 
these sensor-based keyloggers can be improved as our attack allows to infer (1) the exact time when the user touches the screen, and (2) whether the user touches the soft-keyboard or any other region of the display. 

Our attacks only require the user to install a malicious app on the smartphone. However, as shown by Oren~\etal\cite{DBLP:conf/ccs/OrenKSK15}, \PrimeProbe attacks can even be performed  from within browser sandboxes through remote websites using JavaScript on Intel platforms. Gruss~\etal\cite{DBLP:conf/esorics/GrussBM15} showed that JavaScript timing measurements in web browsers on ARM-based smartphones achieve a comparable accuracy as on Intel platforms. Thus, it seems likely that \PrimeProbe through a website works on ARM-based smartphones as well. We expect that such attacks will be demonstrated in future work. The possibility of attacking millions of users shifts the focus of cache attacks to a new range of potential malicious applications.

In our experiments with the predecessor of ART, the 
Dalvik VM, we found that the just-in-time compilation effectively prevents \EvictReload and \FlushReload attacks. The just-in-time compiled code is not shared and thus the requirements for these two attacks are not met.
However, \PrimeProbe attacks work on ART binaries and just-in-time compiled Dalvik VM code likewise.

\section{Attack on Cryptographic Algorithms}\label{sec:crypto}
In this section 
we show how \FlushReload, \EvictReload, and \PrimeProbe can be used to attack AES
T-table implementations that are still in use on Android devices. 
Furthermore, we demonstrate the possibility to infer activities within the ARM TrustZone by observing the cache activity using \PrimeProbe. 
We perform all attacks cross-core and in a synchronized setting, \ie the attacker 
triggers
the execution of cryptographic algorithms by the victim process. Although more sophisticated attacks are possible, our goal is 
to demonstrate that our work enables practical cache 
attacks on smartphones. 

\subsection{AES T-Table Attacks}
Many cache attacks against AES T-table implementations have been demonstrated and appropriate countermeasures have already been proposed. Among these
countermeasures are, \eg so-called bit-sliced implementations~\cite{DBLP:conf/cans/RebeiroSD06,DBLP:conf/ctrsa/Konighofer08,DBLP:conf/ches/KasperS09}.
Furthermore, Intel addressed the problem by adding dedicated instructions for AES~\cite{intel-aes-ni} 
and ARM also follows the same direction with the ARMv8 instruction set~\cite{arm_arch_manualv8}. 
However, our investigations showed that Bouncy Castle, a crypto library widely used in Android apps such as the WhatsApp messenger\cite{Appbrain2016}, still uses a T-table implementation.
Moreover, the OpenSSL library, which is the default crypto provider on recent Android versions, uses T-table implementations until version 1.0.1.\footnote{Later versions use a bit-sliced implementation if ARM NEON is available or dedicated AES instructions if ARMv8-A instructions are available. Otherwise, a T-table implementation is used. This is also the case for Google's BoringSSL library.}
This version is still officially supported and commonly used on Android devices, \eg the \Alcatel.
T-tables contain the precomputed AES round transformations, allowing to perform encryptions and decryptions 
by simple XOR operations. 
For instance, let $p_i$ denote the plaintext bytes, $k_i$ the initial key bytes, and $s_i = p_i \oplus k_i$ the initial state bytes.
The initial state bytes are used to retrieve precomputed T-table elements for the next round. 
If an attacker knows a plaintext byte $p_i$ and the accessed element of the T-table, 
it is possible to recover the key bytes $k_i = s_i \oplus p_i$. However, it is only possible to derive the upper 4 bits of $k_i$ through our cache attack on a device with a cache line size of 64 bytes. 
This way, the attacker can learn 64 key bits. In second-round and last-round attacks the key space can be reduced further. 
For details about the basic attack strategy we refer to the work of Osvik~\etal\cite{DBLP:conf/ctrsa/OsvikST06,DBLP:journals/joc/TromerOS10}.
Although we successfully mounted an \EvictReload attack on the \Alcatel against the OpenSSL AES implementation, we do not provide further insights 
as we are more interested to perform the first cache attack on a Java implementation.

\paragraph{Attack on Bouncy Castle.}

Bouncy Castle is implemented in Java and provides various cryptographic primitives including AES. As Bouncy Castle 1.5 still employs AES T-table implementations by default, all Android devices that use this version are vulnerable to our presented attack. To the best of our knowledge, we are the first to show an attack on a Java implementation.

During the initialization of Bouncy Castle, the T-tables are copied to a local private memory area. Therefore, these copies are 
not shared among different processes. Nevertheless, we demonstrate that \FlushReload and \EvictReload are efficient attacks on such an implementation if shared memory is available. Further, we demonstrate a cross-core \PrimeProbe attack without shared memory that is applicable in a real-world scenario. 

\begin{figure}
  \centering
  \includegraphics[width=0.49\hsize]{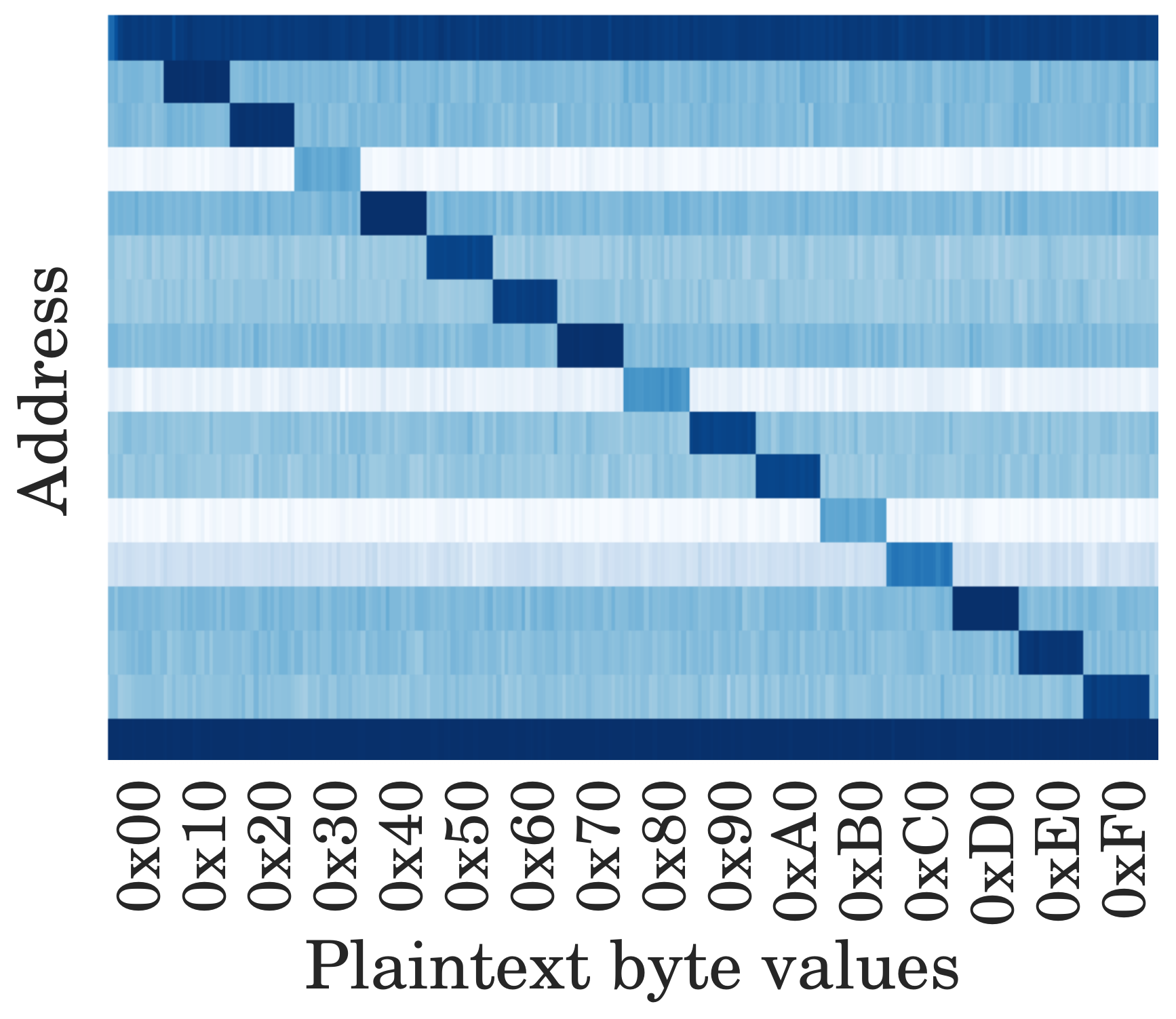}
  \hfill
  \includegraphics[width=0.49\hsize]{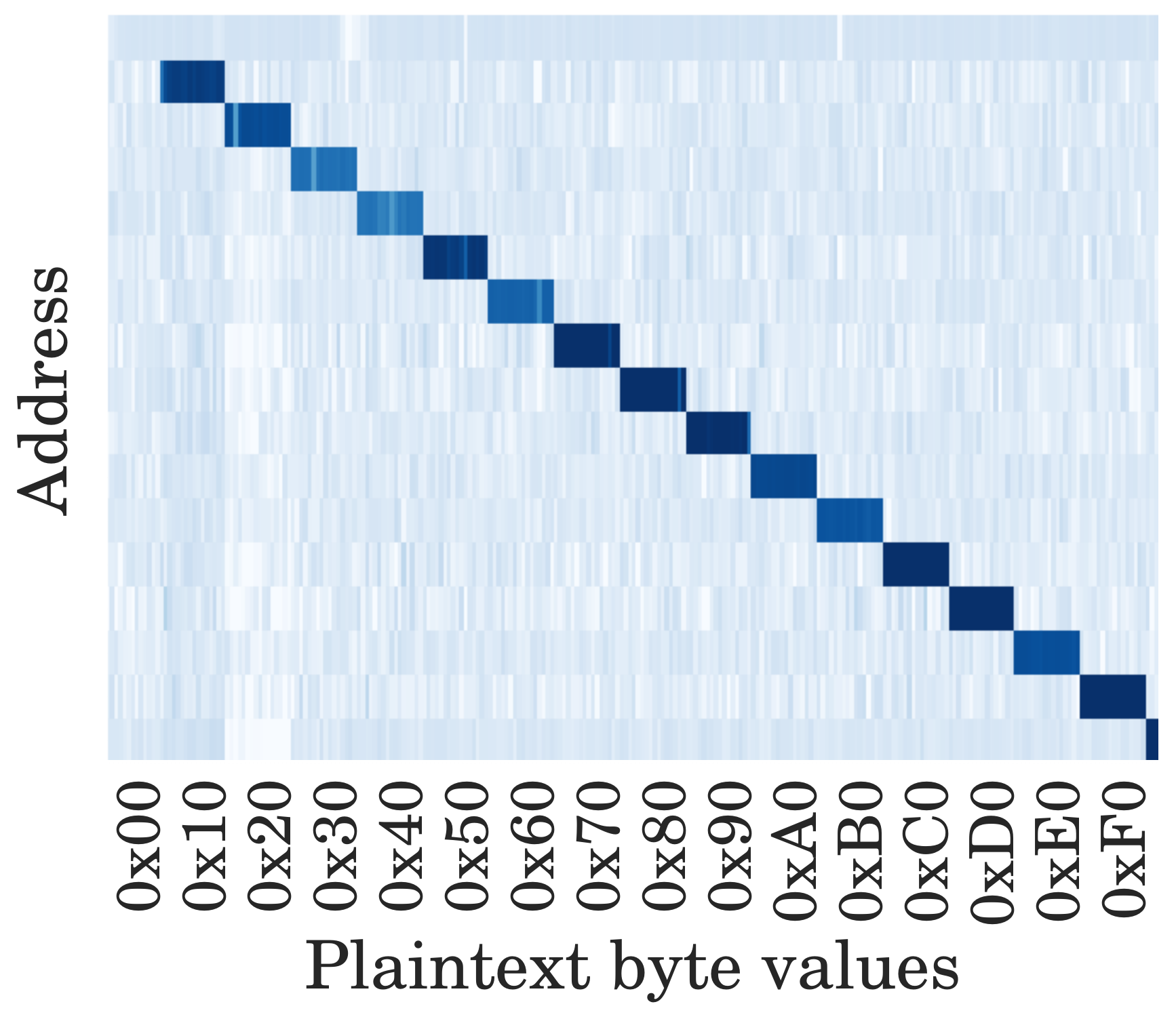}
  
  \caption{Attack on Bouncy Castle's AES using \EvictReload on the \Alcatel (left) and \FlushReload on the \Samsung (right).} 
  \label{fig:aes_bouncy_castle_e_r}
\end{figure}

Figure~\ref{fig:aes_bouncy_castle_e_r} shows a
template matrix of the first T-table for all 256 values for plaintext byte $p_0$ and a key that is fixed 
to $0$ while the remaining plaintext bytes are random.
These plots reveal the upper 4 key bits of $k_0$~\cite{DBLP:conf/ctrsa/OsvikST06,DBLP:conf/cosade/SpreitzerP13}. Thus, in our case the key space is reduced to 64 bits after 256--512 encryptions. We consider a first-round attack only, because we aim to demonstrate the applicability of these attacks on ARM-based mobile devices. However, full-key recovery is possible with the same 
techniques by considering more sophisticated attacks targeting different rounds~\cite{DBLP:journals/joc/TromerOS10,DBLP:journals/cj/SavasY15}, even for asynchronous attackers~\cite{DBLP:conf/raid/ApececheaIES14,DBLP:conf/cosade/GulmezogluIAES15}.

We can exploit the fact that the 
T-tables are placed on a different boundary every time the process is started.
By restarting the victim application we can obtain arbitrary disalignments of T-tables. 
Disaligned T-tables allow to reduce the key space to 20 bits on average and for specific disalignments even full-key recovery without a single brute-force computation is possible~\cite{DBLP:conf/cosade/SpreitzerP13,DBLP:conf/acisp/TakahashiFAF13}.
We observed not a single case where the T-tables were aligned. Based on the first-round attack matrix in Figure~\ref{fig:aes_bouncy_castle_e_r}, the expected number of encryptions until a key byte is identified is $1.81 \cdot 128$. Thus, full key recovery is possible after $1.81 \cdot 128 \cdot 16 = 3\,707$ encryptions by monitoring a single address during each encryption.

\begin{figure}
\centering
\begin{tikzpicture}
\begin{axis}[
style={font=\footnotesize},
xlabel=Execution time in CPU cycles,
ylabel=Number of cases,
width=0.95\hsize,
ymin=0,
height=4cm
]
\addplot+[no marks] table[x=Time,y=Hit] {hit_vs_miss_alto45_direct_pp.csv};
\addlegendentry{Victim access}
\addplot+[no marks,densely dashed] table[x=Time,y=Miss] {hit_vs_miss_alto45_direct_pp.csv};
\addlegendentry{No victim access}
\end{axis}
\end{tikzpicture}
  \caption{Histogram of \PrimeProbe timings depending on whether the victim accesses congruent memory on the ARM Cortex-A53.} 
  \label{fig:histogram_hit_miss_pp}
\end{figure}
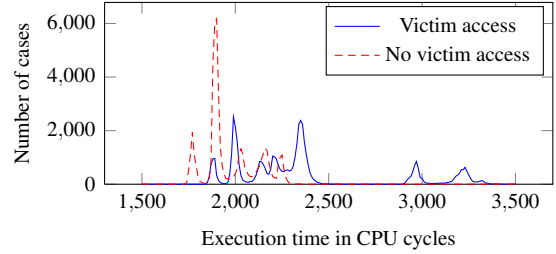

\paragraph{Real-world cross-core attack on Bouncy Castle.}
If the attacker has no way to share a targeted memory region with the victim, 
\PrimeProbe instead of \EvictReload or \FlushReload can be used. This is the case for dynamically generated data or private memory of another process. 
Figure~\ref{fig:histogram_hit_miss_pp} shows the \PrimeProbe histogram for cache hits and cache misses. We observe a higher execution time if the victim accesses a congruent memory location. Thus, \PrimeProbe can be used for a real-world cross-core attack on Bouncy Castle and also allows 
to exploit disaligned T-tables as mentioned above. 

In a preprocessing step, the attacker identifies the cache sets to be attacked by performing random encryptions and searching for active cache sets. 
Recall that the cache set (index) is derived directly from the physical address on ARM, 
\ie the lowest $n$ bits determine the offset within a $2^n$-byte cache line and the next $s$ bits determine one of the $2^s$ cache sets.
Thus, we only have to find a few cache sets where a T-table maps to in order to identify all cache sets required for the attack. 
On x86 the replacement policy facilitates this attack and allows 
even to deduce the number of ways that have been replaced in a specific cache set~\cite{DBLP:conf/ctrsa/OsvikST06}. On ARM the random replacement policy makes \PrimeProbe more difficult as cache lines are replaced in a less predictable way.
To launch a \PrimeProbe attack, we apply the eviction strategy and the crafted reaccess patterns we described in Section~\ref{sec:eviction}.

\begin{figure}
  \centering
  \includegraphics[width=0.8\hsize]{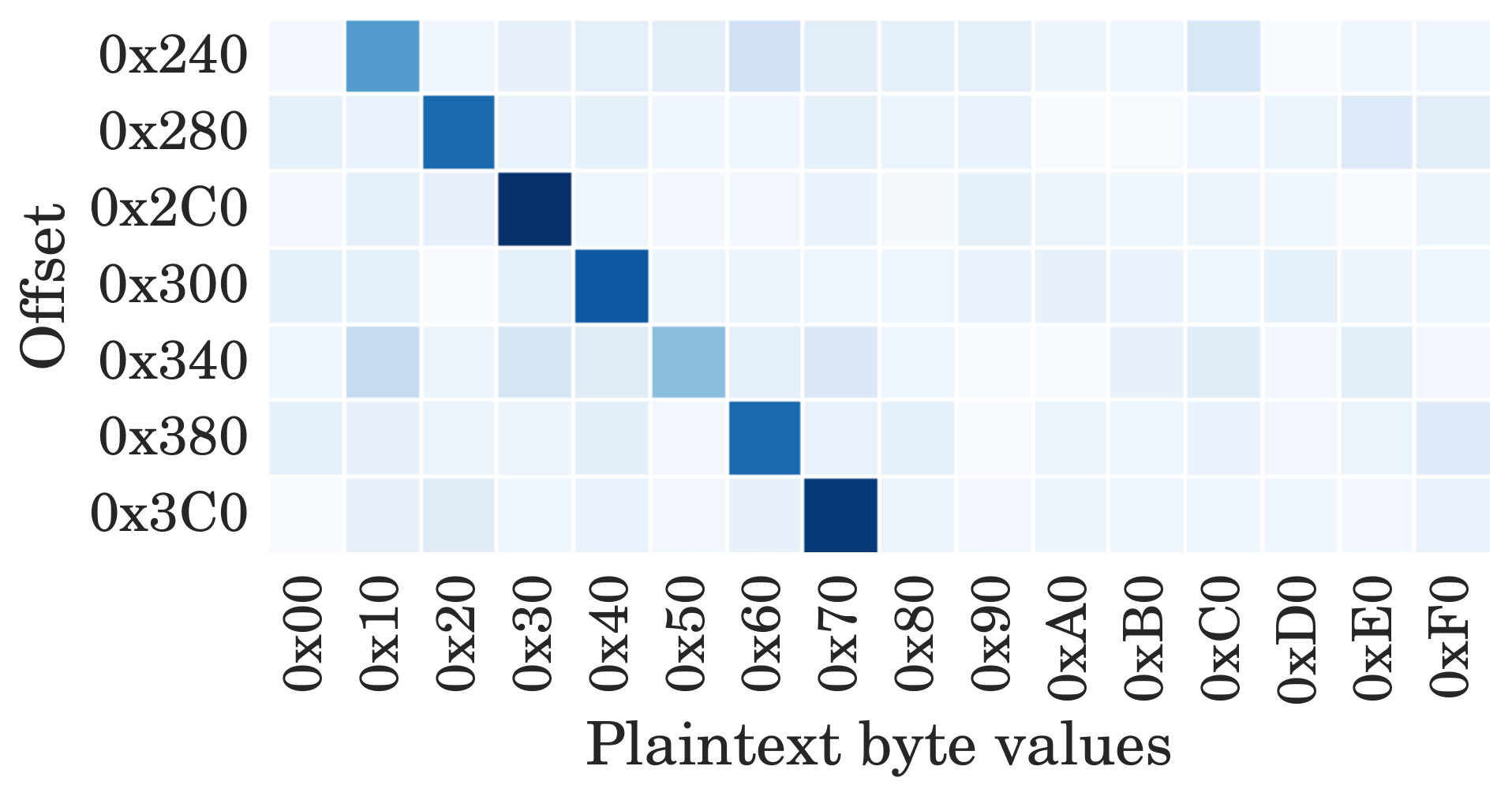}

  \caption{Excerpt of the attack on Bouncy Castle's AES using \PrimeProbe.} 
  \label{fig:aes_bouncy_castle_p_p}
\end{figure}

Figure~\ref{fig:aes_bouncy_castle_p_p} shows an excerpt of the cache template matrix resulting from a \PrimeProbe attack on one T-table. For each combination of plaintext byte and offset we performed $100\,000$ encryptions for illustration purposes. We only need to monitor a single address to obtain the upper 4 bits of $s_i$ and, thus, 
the upper 4 bits of $k_i = s_i \oplus p_i$. 
Compared to the \EvictReload attack from the previous section, 
\PrimeProbe requires 3 times as many measurements to achieve the same accuracy. Nevertheless, our results show that an attacker can run \PrimeProbe attacks on ARM CPUs just as on Intel CPUs.

\subsection{Spy on TrustZone Code Execution}
The ARM TrustZone is a hardware-based security technology built into 
ARM CPUs to provide a secure execution environment~\cite{arm_arch_manualv8}.
This 
trusted execution environment is isolated from the \textit{normal world} using hardware support.
The TrustZone is used, \eg as a hardware-backed credential store, to emulate secure elements for payment
applications, digital rights management as well as verified boot and kernel integrity measurements. 
The services are provided by so-called trustlets, \ie applications that run in the secure world. 

Since the secure monitor can only be called from the supervisor context, the
kernel provides an interface for the userspace to interact with the
TrustZone. On the \Alcatel, the TrustZone is
accessible through a device driver called QSEECOM (Qualcomm Secure Execution
Environment Communication) and a library \texttt{libQSEEComAPI.so}. The key master trustlet on the \Alcatel provides an interface to generate hardware-backed RSA keys, which can then be used inside the TrustZone to sign and verify signatures. 

Our observations showed that a \PrimeProbe attack on the TrustZone is not much different from a \PrimeProbe attack on any application in the normal world.
However, as we do not have access to the source code of the TrustZone OS or any trustlet, we only conduct simple attacks.\footnote{More sophisticated attacks would be possible by reverse engineering these trustlets.}
We show that \PrimeProbe can be used to distinguish whether a provided key is valid or not. 
While this might also be observable through the overall execution time, we demonstrate
that the TrustZone isolation does not protect against cache attacks from the normal world and any trustlet can be attacked.

\begin{figure}[t!]
\centering
\begin{tikzpicture}
\begin{axis}[
style={font=\footnotesize},
xlabel=Set number,
ylabel=Probing time in CPU cycles,
width=\hsize,
xmin=250,
xmax=350,
height=4cm
]
\addplot+[no marks,blue, densely dotted] table[x=Set,y=Valid1] {tz_pp_alto45_mse.csv};
\addlegendentry{Valid key 1}
\addplot+[no marks, red, dashed] table[x=Set,y=Valid2] {tz_pp_alto45_mse.csv};
\addlegendentry{Valid key 2}
\addplot+[no marks, green, dashed] table[x=Set,y=Valid3] {tz_pp_alto45_mse.csv};
\addlegendentry{Valid key 3}
\addplot+[no marks, black, solid] table[x=Set,y=Invalid] {tz_pp_alto45_mse.csv};
\addlegendentry{Invalid key}
\end{axis}
\end{tikzpicture}
  \caption{Mean squared error between the average \PrimeProbe timings of valid keys and invalid keys on the \Alcatel.}
  \label{fig:histogram_tz_alto45}
\end{figure}
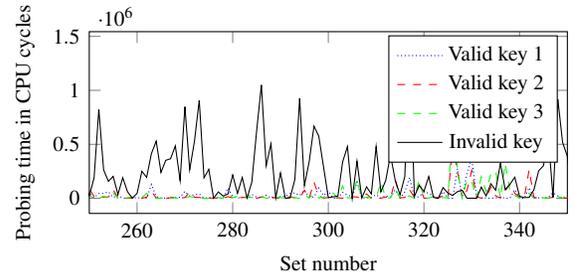

We evaluated cache profiles for multiple valid as well as invalid keys. Figure~\ref{fig:histogram_tz_alto45} shows the mean squared error over two runs for different valid keys and one invalid key compared to the average of valid keys. We performed \PrimeProbe before and after the invocation of the corresponding trustlet, \ie prime before the invocation and probe afterwards. 
We clearly see a difference in some sets (cache sets 250--320) that are used during the signature
generation using a valid key.
These cache profiles are reproducible and can be used to distinguish whether a valid or an invalid key has 
been used in the TrustZone. Thus, the secure world leaks information to the non-secure world.

On the \Samsung, the TrustZone flushes the cache when entering or leaving the trusted world. However, by performing a \PrimeProbe attack in parallel, \ie multiple times while the trustlet performs the corresponding computations, the same attack can be mounted.

\section{Countermeasures}\label{sec:countermeasures}
Although our attacks exploit hardware weaknesses, software-based countermeasures 
could impede such attacks. Indeed, we use 
unprotected access to system information that is available on all Android versions. 

As we have shown, the operating system cannot prevent access to timing information. However, other information supplied by the operating system that facilitates these attacks could be restricted. For instance, we use \verb|/proc/pid/| to retrieve information about any other process on the device, \eg \verb|/proc/pid/pagemap| is used to resolve virtual addresses to physical addresses. Even though access to \verb|/proc/pid/pagemap| and \verb|/proc/self/pagemap| has been restricted in Linux in early 2015, the Android kernel still allows access to these resources. Given the immediately applicable attacks we presented, we stress the urgency to merge the corresponding patches into the Android kernel. 
Furthermore, we use \verb|/proc/pid/maps| to determine shared objects that are mapped into the address space of a victim. Restricting access to procfs to specific privileges or permissions would make attacks harder. We recommend this for both the Linux kernel as well as Android. 

We also exploit the fact that access to shared libraries as well as \texttt{dex} and \texttt{art} optimized program binaries 
is only partially restricted on the file system level. While we cannot retrieve a directory listing of \verb|/data/dalvik-cache/|, all files are readable for any process or Android application. 
We recommend to allow read access to these files to their respective owner exclusively
to prevent \EvictReload, \FlushReload, and \FlushOnly attacks through these shared files.

In order to prevent cache attacks against AES T-tables, hardware instructions should be used. If this is not an option, a software-only bit-sliced implementation must be employed, especially when disalignment is possible, as it is the case in Java. Since OpenSSL 1.0.2 a bit-sliced implementation is available for devices capable of the ARM NEON instruction set and dedicated AES instructions are used on ARMv8-A devices. Cryptographic algorithms can also be protected using cache partitioning~\cite{DBLP:conf/hpca/LiuGYMRHL16}. However, cache partitioning comes with a performance impact and it can not prevent all attacks, as the number of cache partitions is limited.

We responsibly disclosed our attacks and the proposed countermeasures to Google and other development groups prior to the publication of our attacks.
Google has applied upstream patches preventing access to \verb|/proc/pid/pagemap| in early 2016 and recommended installing the security update in March 2016~\cite{Android2016March}.

\section{Conclusion}\label{sec:conclusion}
In this work we demonstrated the most powerful cross-core cache attacks \PrimeProbe, \FlushReload, \EvictReload, and \FlushOnly on default configured unmodified Android smartphones. Furthermore, these attacks do not require any permission or privileges.
In order to enable these attacks in real-world scenarios, we have systematically solved all challenges that prevented highly accurate cache attacks on ARM so far.
Our attacks are the first cross-core and cross-CPU attacks on ARM CPUs. Furthermore, our attack techniques provide a high resolution and a high accuracy, which allows monitoring singular events such as 
touch and swipe actions on the screen, touch actions on the soft-keyboard, and inter-keystroke timings.
In addition, we show that efficient state-of-the-art key-recovery attacks can be
mounted against the default AES implementation that is part of the Java Bouncy
Castle crypto provider and that cache activity in the ARM TrustZone can be
monitored from the normal world.

The presented example attacks are by no means exhaustive and launching our proposed attack against other libraries and apps will reveal numerous further exploitable information leaks.
Our attacks are applicable to hundreds of millions of today's off-the-shelf smartphones as they all have very similar if not identical hardware.
This is especially daunting since smartphones have become the most important personal computing devices and our techniques significantly broaden the scope and impact of cache attacks. 

\section*{Acknowledgment}
We would like to thank our anonymous reviewers for their valuable comments and suggestions.

\noindent\begin{tabular}{m{\dimexpr 0.13\hsize} m{2pt} m{\dimexpr 0.87\hsize-5\tabcolsep-2pt}}
\vspace*{1mm}\includegraphics[width=\hsize]{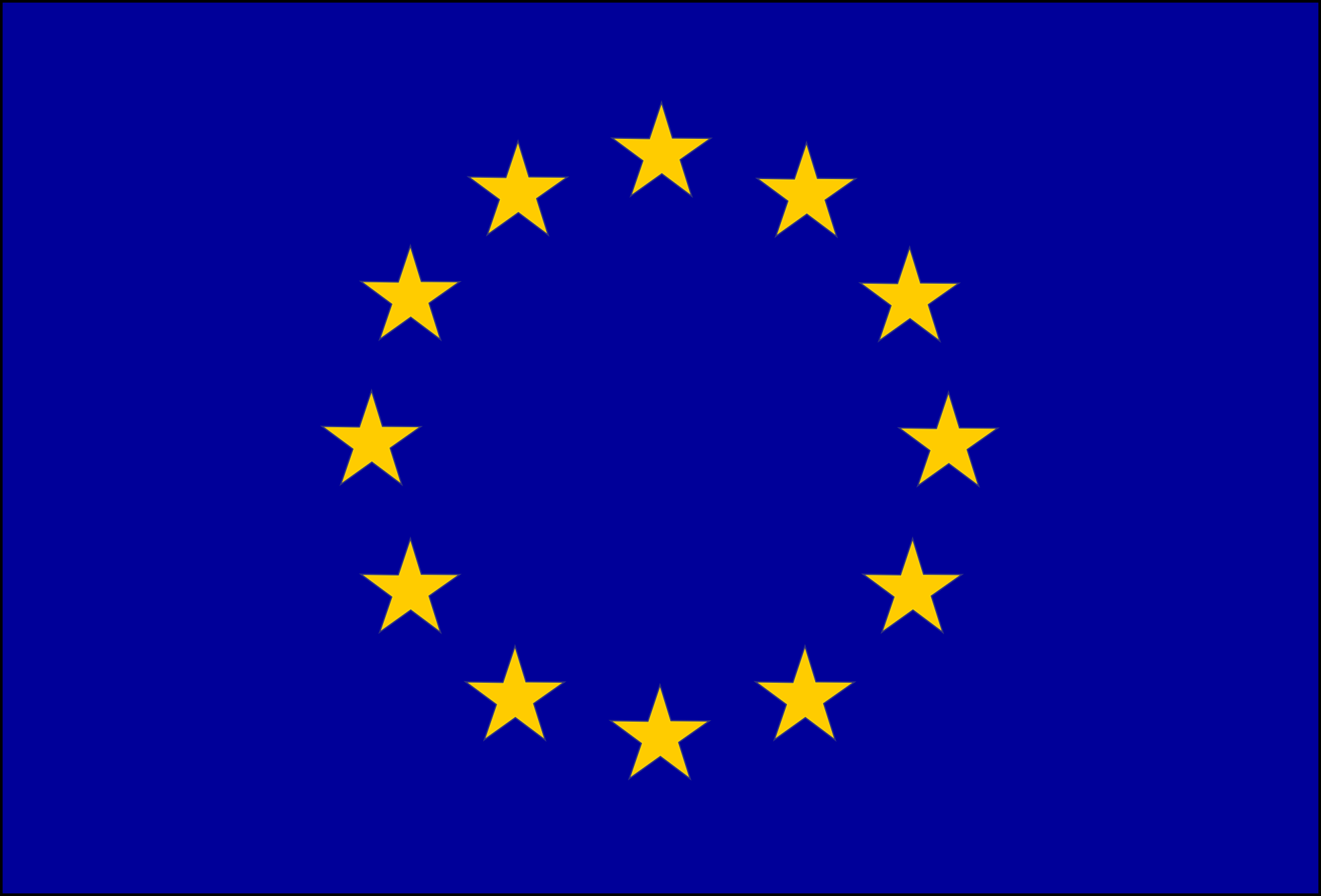}& &Supported by the EU Horizon 2020 programme under GA No. 644052 (HECTOR), the EU FP7 programme under GA No. 610436 (MATTHEW), and the Aus-\quad \quad \quad \vspace{-\baselineskip}
\end{tabular} \\
trian Research Promotion Agency (FFG) under grant number 845579 (MEMSEC).

{\footnotesize \bibliographystyle{acm}
\bibliography{bibliography}}

\end{document}